# An Enhanced Binary Particle-Swarm Optimization (E-BPSO) Algorithm for Service Placement in Hybrid Cloud Platforms

Wissem ABBES[(a)(b)], Zied KECHAOU[(b)], Amir HUSSAIN[(c)], Abdulrahman M. QAHTANI[(d)], Omar AlMUTIRY [(e)], Habib DHAHRI [(e)], Adel M. ALIMI[(b)(f)]

*(a) University of Sousse, ISITCom, 4011, Sousse, Tunisia*
*(b) University of Sfax, National Engineering School of Sfax (ENIS), REGIM-Lab.: REsearch Groups on Intelligent Machines, BP 1173, 3038, Sfax, Tunisia*
*(c) School of Computing, Edinburgh Napier University, UK*
*(d) Department of Computer Science, College of Computers and Information Technology, Taif University, P.O. Box. 11099, Taif 21944, Saudi Arabia.*
*(e) College of Applied Computer Science, King Saud University, (Almuzahmiyah Campus), Riyadh, Saudi Arabia*
*(f) Department of Electrical and Electronic Engineering Science, Faculty of Engineering and the Built Environment, University of Johannesburg, South Africa.*
Corresponding author: wissem.abbes@enis.usf.tn

---

**Abstract**
Nowadays, hybrid cloud platforms stand as an attractive solution for organizations intending to implement combined private and public cloud applications, in order to meet their profitability requirements. However, this can only be achieved through the utilization of available resources while speeding up execution processes. Accordingly, deploying new applications entails dedicating some of these processes to a private cloud solution, while allocating others to the public cloud. In this context, the present work is set to help minimize relevant costs and deliver effective choices for an optimal service placement solution within minimal execution time. Several evolutionary algorithms have been applied to solve the service placement problem and are used when dealing with complex solution spaces to provide an optimal placement and often produce a short execution time.

The standard BPSO algorithm is found to display a significant disadvantage, namely, of easily trapping into local optima, in addition to demonstrating a noticeable lack of robustness in dealing with service placement problems. Hence, to overcome critical shortcomings associated with the standard BPSO, an Enhanced Binary Particle Swarm Optimization (E-BPSO) algorithm is proposed, consisting of a modification of the particle position updating equation, initially inspired from the continuous PSO.
Our proposed E-BPSO algorithm is shown to outperform state-of-the-art approaches in terms of both cost and execution time, using a real benchmark.

*Keywords:* Placement Optimization, Particle Swarm Optimization, E-BPSO Algorithm, Hybrid Cloud, Service Based Application.

## 1. INTRODUCTION

With the advent of cloud computing technologies, a growing number of content distribution applications have started to veer to cloud-based services for the sake of maintaining relatively high-level scalability and effective cost rates. Actually, cloud systems turn out to be indispensable technological tools, versatile for the persistence of modern society.

Hence, for a company's applications to be effectively deployed on a private cloud basis, it is necessary that resources be provided, managed, and maintained via its proper infrastructure. Noteworthy, however, is that should a company decide to incorporate other applications fit for implementation with its proper private cloud infrastructure, it turns out to be under the constraint of using extra outside resources or a public cloud. This highlights well that the private cloud system associated with physical limitations needs to be entirely accounted for rather effectively. Such a situation appears to occur mainly when applications tend to opt for further intensification procedures, entailing additional resources, which private cloud cannot provide, or when a newly deployed application cannot be exclusively satisfied through a private Cloud. Hence, acquiring extra resources available in a public cloud might well help in handling such situations. Therefore, companies are increasingly appealing to cloud-based environments to extend the deployment and running range of the available applications.

Regarding the present work scope, it is primarily focused on investigating service placement optimization within a hybrid-cloud context. The cloud platforms, subject of study, involve business service solutions providing units or components, increasingly used to implement service-based applications. Once service-based applications are implemented into a hybrid cloud, selecting the components that fit public clouds remains an open-ended question. However, it is worth noting that several parameters need to be considered when deciding on the appropriate application components subject to a transfer to the public cloud, mainly privacy, QoS, security, communication costs, and the hosting costs relocation. Similarly, the execution time criterion stands as a critical parameter.

The present work's placement problem consists mainly of the NP-Hard nature of the problem [1], which relates mainly to the non-availability of optimal algorithms enabling us to solve issues as promptly and effectively as possible. Several algorithms and models from the computational intelligence field have been applied to solve the service placement problem, i.e., Genetic algorithm (GA) [2], Ant Colony Optimization (ACO) [3], heuristic-based [4], PSO [5]. Heuristics have been implemented to resolve the cloud service placement problem in which immediate solutions cannot be determined due to the hugeness of the cloud environment and the high number of services. Consequently, the present work is conceived to treat each service component separately, i.e., to be deployed either as part of the private cloud or within the public cloud context. Accordingly, the suitable placement service problem is essentially a binary optimization problem. Hence, the purpose of conducting the present research is to minimize the public cloud incurred costs while preserving a minimum execution time. These two criteria constitute essential factors for a company to persist successfully in a competitive world. To this end, an Enhanced BPSO algorithm is put forward as an adequately effective service placement solution, useful for application within a hybrid cloud context. Accordingly, each particle's respective position would be equal either to zero (if the service is liable to be deployed in a private cloud) or to one (if it proves to be applicable within the public cloud context).

Thus, the advanced E-BPSO undertakes to substitute the sigmoid function by our proposed equation to help resolve the problem of premature convergence and local optima, significantly affecting the service placement application in hybrid cloud, which makes it highly distinguishable and relevant to the standard BPSO algorithm.

The remainder of this paper is organized as follows. Section 2 involves a formulation of the Service-Based Application (SBA) placement problem conceived within a hybrid-cloud context, along with a graphic illustration of a SBA scheme. Section 3 is devoted to depicting the major relevant works dealing with cloud data placement strategy, highlighting the major contribution and achievement brought about by each work, compared to the present research modest contribution and reached findings. Concerning Section 4, it englobes a presentation of our proposed E-BPSO algorithm, envisioned to deal with the placement strategy issue. The implemented experimental analysis and evaluation procedures are dealt with in section 5, while the ultimate section depicts the conclusions and perspective.

## 2. PROBLEM FORMULATION

### *2.1. Service placement problem in hybrid cloud*

It is worth recalling that the Service-Based Application (SBA) in the hybrid cloud has been the subject of our previously designed framework, subject of our already published work referenced as [6], which we dubbed as a hybrid composition of private and public cloud structures. SBA is a set of basic services to provide flexibility to complex environments, widely scattered by dispersed ranges and arrays of services each environment maintains.

It is natural for any organization that the cloud applications should be deployed privately, as long as the private cloud can provide the required resource needs. Yet, an appeal could be made to SBA to be deployed via public cloud mainly: (first) when the deployed applications are discovered to require greater resources that private cloud could not provide; (second) when private cloud proves to demonstrate insufficiency to satisfy a newly applied deployment request; or (third) when the utilized applications release resources indicating that a new deployment procedure must be implemented to release allocated resources from the public cloud withheld data.

In this respect, we consider setting up a specifying threshold to determine the appropriate case for making a rightful decision to appeal to the public cloud. This threshold can be quantified via hosting units, hence the notation HQ that designates a hosting quantity. Accordingly, the quantity of resources required for executing any selected services has to be greater or equal to HQ range, as indicated below.

At this level, we consider applying the same basic problem-formulation definition as that used in [7], specifically:

$$Minimize(H + PC + HC) \quad Subject_to: \tag{1}$$

$$\sum_{i=1}^{n} h(s_i) \times l(s_i) \geq HQ \quad Where: \tag{2}$$

$$H = \sum_{i=1}^{n} \alpha \times h(s_i) \times l(s_i) \tag{3}$$

$$PC = \sum_{e=<s_i,s_j>\in E} \beta_2 \times c(e) \times l(s_i) \times l(s_j) \tag{4}$$

$$HC = \sum_{i=1}^{n} \sum_{j(s.t.(e=<s_i,s_j>\in E))} \beta_1 \times c(e) \times l(s_i) \times (1 - l(s_j)) \tag{5}$$

Where: (1) stands for the objective function (minimize H (Hosting cost) + PC (Public Communication cost) + HC (Hybrid Communication cost));

(2): designates a constraint equation that represents the sum of hosting quantity (HQ) of the public cloud deployed services, which has to be greater than or equal to HQ (minimum threshold). HQ is a value assigned by a resource request case. The need for the public cloud resource can be quantified in terms of the amount of hosting units (units of platform resources required). Once a public cloud request is triggered, one has to decide on the appropriate application services to opt for, to be deployed as part of public cloud-based service. In this case, the quantity of required platform resources necessary for providing the selected services has to be greater than, or equal to, the HQ associated ones;

(3): is the sum of hosting service costs deployed in the public cloud;

(4): designates the sum of publicly made communications (established between the public cloud deployed services);

and (5): is the sum of hybrid sustained communications (maintained by the public cloud deployed services and those ensured by private cloud).

Table 2 depicts the various abbreviated designations used in the problem formulation stage.

**Table 2.** The problem formulation abbreviations and relevant designations

| | |
|---|---|
| **h()** | is a hosting function that associates a positive number to each service representing its hosting quantity of needed resources for its deployment. |
| **c()** | is a communication function that associates to each edge $e = < s_i, s_j > \in E$ a positive number representing the communication rate established between $s_i$ and $s_j$. |
| **l()** | is a location function that for each service associates $0$ if it is deployed in the private cloud and $1$ if it is deployed in the public cloud. |
| **α** | is the cost of a resource hosting unit. |
| **β1** | is the cost of a communication unit between the public and the private cloud. |
| **β2** | is the cost of a unit of communication as established inside the public cloud. |
| **H** | is the sum of hosting costs of services as deployed within the public cloud. Indeed, the expression $\alpha \times h(s_i) \times l(s_i)$ (where α is the cost of a resource hosting unit, $h(s_i)$ is hosting quantity of service $s_i$ and $l(s_i)$ takes the value $1$ if $s_i$ relates to the public cloud, and $0$ otherwise) which is equal to the $s_i$ hosting cost if this service is maintained within the public cloud. |
| **PC** | is the sum of publicly established communications (communications between services deployed within the public cloud). Indeed, there is a public communication maintained between $s_i$ and $s_j$ if they are both deployed in the public cloud. $l(s_i) \times l(s_j) = 1$. |
| **HC** | stands for the sum of hybrid communications (communications between services deployed in the public cloud and those deployed in the private cloud). Indeed, there is a hybrid communication between $s_i$ and $s_j$ if one of them is deployed in the public cloud and the other one is deployed in the private cloud, since, either $l(s_i) \times (1 - l(s_j)) = 1$ or $l(s_j) \times (1 - l(s_i)) = 1$. |

In addition to minimizing the hosting and communication costs, we are also interested in reducing the execution time. In effect, execution time represents an important criterion, whereby the decision-taking process can be executed promptly, thus, ensuring the company significant savings.

*2.2. Case study*

In this context, a structured bank account opening process [8] is illustrated in the below figuring Business Process Model and Notation (BPMN [9]) diagram.

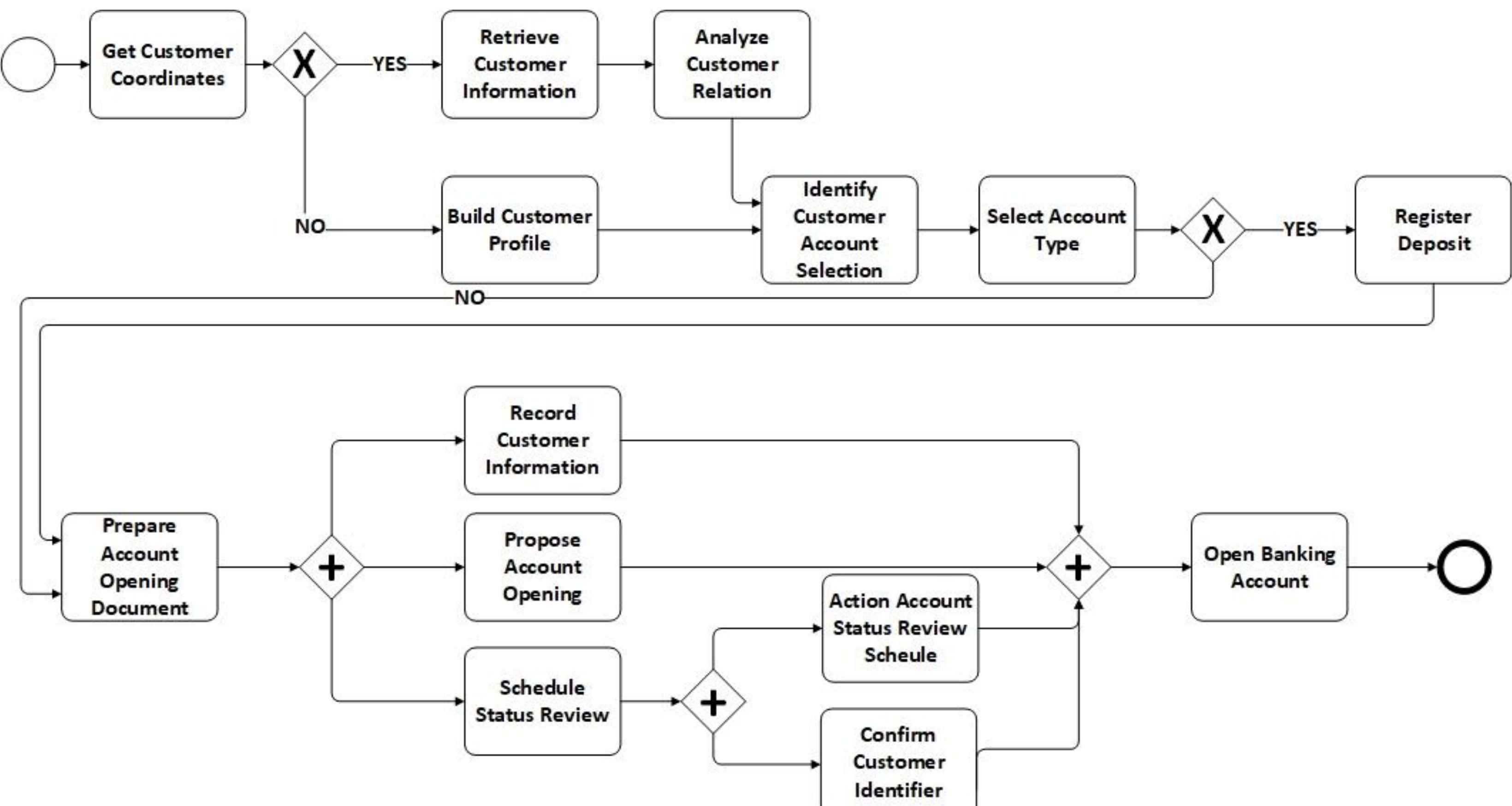

**Figure 1:** A sample SBA application modeled following the structured process.

The SBA, depicted in Figure 1, can also be partially modelled in the form of a graph, as appearing in Figure 2, where services and gateway nodes are represented by graph nodes and inter-service connections/transitions by edges. Nodes are identified by the number of resource hosting units and characterized by the corresponding quantified amount. Edges are characterized by an amount of established communication transactions, referring to the amount of traffic transferred onto the considered corresponding edge.

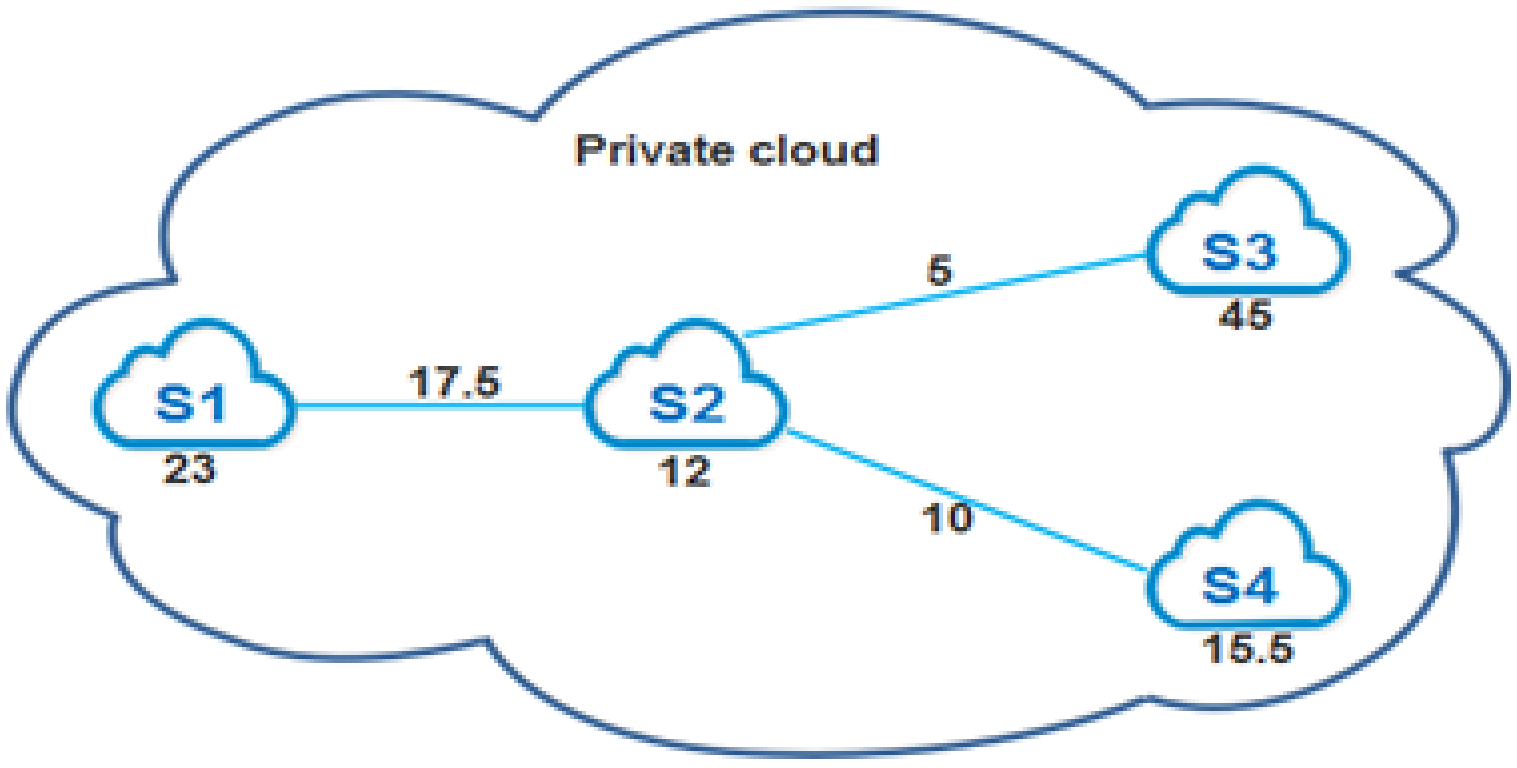

**Figure 2:** An SBA graph corresponding to opening a bank account.

The SBA graph, as illustrated in Figure 2, appears to enclose distinct services. Each service incorporates a hosting quantity, and each edge englobes a quantity of communication unit.

As deployed in a hybrid cloud environment, the SBA is represented by a graph, where some services are maintained via public cloud, while others are ensured via private cloud, as shown in Figure 3 below.

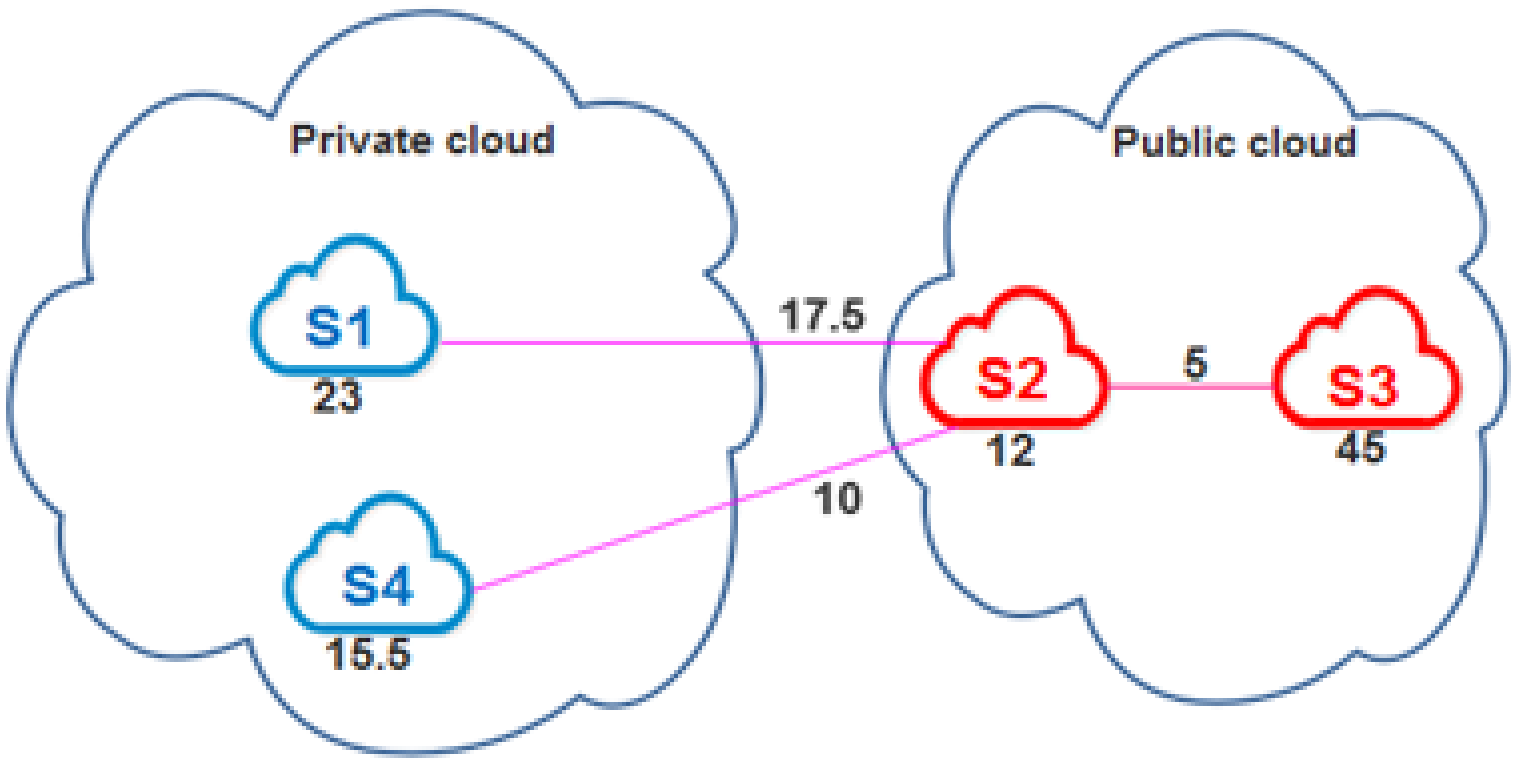

**Figure 3:** Bank account opening process as deployed within a hybrid cloud architecture.

Accordingly, inter-cloud communication costs depend highly on the services' placement mode (i.e., between private and public clouds). Indeed, the cost of intra-cloud maintained communication (exclusively within public clouds) appears to differ remarkably from the inter-cloud maintained one. It is worth noting, at this level, that both communication and hosting costs are not considered within the private cloud framework.

The model applications' deployment cost, as depicted in Figure 3, for $\alpha$=40, $\beta_1$=20, $\beta_2$=10, is determined as follows: $40\times(12+45) + 20\times(17.5+10) + 10\times5 = 2880$. Thus, no communication and hosting cost considerations have been accounted for regarding the private cloud mode case, as we assume that a company maintains its proper private cloud system. Therefore, these costs will not be computed since these services, and the relating costs, are generated on a private cloud basis.

## 3. RELATED WORKS

The resource placement problem (virtual machines, Web servers, software components and data), remarkably persistent in cloud environments, has been addressed from different perspectives. More specifically, it has been approached through accounting for the different cloud model versions available (private, public, hybrid and federated Clouds) or by considering the various associated criteria (hosting, communication, QoS, etc.)

Generally, the type of cloud application most often opted for is either hybrid or federated, wherein the relevant infrastructure involves two or more cloud models (public or private) operating independently. In effect, the two predominant cloud modes turn out to be the hybrid cloud and the federated cloud. Regarding the present work, the adopted hybrid cloud definition follows that set by Van den Bossche et al. [10] and Luong et al. [11]. Accordingly, organizations appeal to public clouds for the sake of meeting their demand and urgent needs for computational resources that exceed their private cloud available capacity. As to federated clouds [12], technology is needed to maintain a joint combination of disparate public clouds, including those owned by different organizations. Indeed, it is only through federation (including the relevant interoperability requirement) can a single cloud provider take advantage of the aggregated capabilities available to provide a seemingly infinite computing service utility.

The most recently developed approach, useful for modelling and computing resource management optimization problems, is the Service placement approach, oriented to cloud users intending to optimize their services' costs and performance. In this respect, the service placement problem provides several responses to the users' needs in Service as a Service (SaaS) and Business Process as a Service (BPaaS). It provides solutions relevant to the optimal allocation of any complex service onto an available set of (virtualized) resources. The nature of resources may vary from physical infrastructure to Virtual Machines (VM) and running software components. Given the fact that service and business processes are commonly defined as constant workflows of tasks or activities [13-15], the standard objectives usually tend to lie in catering for the cost and QoS dimensions, referred to as service throughput [14] or latency [16-18].

In this regard, a well-determined selection of state-of-the-art works dealing with the subject of service placement in private/public, and hybrid/federated clouds, is depicted. Each work should represent a specific helpful approach for deciding on a specific service package while accounting for some crucial criteria or parameters considered in these approaches.

### *3.1. Service placement in private/public cloud*

Huang et al. [20] defined two graph structures: a Service Dependency Graph (SDG), which reflects the inter-service communication cost and execution time of each service, and a Service Concurrency Graph (SCG), which describes possible parallel services. These graphs allowed interdependent services to be co-deployed on the same VM, based on the k-cut principle, where k denotes the number of hosting VM solvers. For Green Monster [17], the focus of interest is laid on a multi-objective Evolutionary Algorithm (EA) using a local search to help optimize Renewable Energy consumption (RE), Cooling Energy consumption (CE) and User-to-Service Distance (USD). Green Monster achieved results that proved to outperform static and random placement provided performance. Also, Yao et al. [18] propose a scheduling algorithm for searching for the optimal service instance allocation, based on Simulated Annealing (SA) and (GA). The objective is to minimize computing power and network latency. Concerning Yusoh and Tang, they document a series of studies [13], [21], [22] that elaborated on the issue of SaaS component placement to help in catering predominantly for the problem of service components mapping to VMs and storage systems. The proposed solutions include a Cooperative Co-evolutionary Genetic Algorithm (CCGA) [21] that splits or dislocates populations into groups to optimize the computation and storage of allocations. In turn, Wada et al. [14] advanced an optimization framework, dubbed E3, to solve the SLA-aware Service Composition (SSC) problem. The E3 framework enables to define a service composition model by putting forward two heuristic algorithms, labelled E3 Multi-Objectives GA (E3-MOGA) and Extreme-E3 (X-E3), designed to solve the SSC related problem. Similar objectives are targeted by SanGA [16], whereby a Self-adaptive network-aware GA is conceived to deal with the service composition associated problem. Worth recalling, in this respect, is that reducing latency and minimizing cost prove to stand as significant optimization targeted objectives. To this end, Li et al. [15] put forward a higher cloud computing layer BPaaS.

Additionally, they envisage resolving the service placement problem frequently persistent within the cloud logistics domain through the implementation of the PSO. In effect, the optimization process's overall objective is to enclose a weighted sum of time, price, availability, and reliability. With respect to Filelis-Papadopoulos et al. [4], they devised a particular simulation framework for implementation with virtual Content Distribution Networks (vCDN). Their framework involves a special optimization scheme, whereby simulation outputs are used to guide the search for optimal cache placements. For AntPu [23], a meta-heuristic approach is proposed for placing VMs to minimize energy consumption and SLA violations in the cloud datacenter. In [24], the authors adopted a cooperative learning strategy while considering security as an additional placement constraint. The authors started by decomposing a population of computing and storage servers into a set of cooperating sub-swarms. The idea is to evaluate each sub-swarm and influence the worst placement plans by the best ones. In this way, the worst placement plans are improved by learning from the best servers in the associated sub-swarms. The survival levels determine the choice of the most secure hosting servers. Multi-population PSO variants have been used in the context of SaaS placement. The composite particles in [5] were applied to represent the entire placement scheme, i.e., a set of hosting virtual machines for a composite SaaS, while sub-swarms in [25] defined candidate compute/storage servers to host a component application/data of an involved SaaS. Their goal was to reduce the SaaS execution time and resource utilization.

*3.2. Service placement in hybrid cloud*

More recently, a significant number of conducted studies have attempted to address several aspects of the service placement issue within the hybrid cloud context. Worth citing among them is the study conducted by Kaviani et al. [26], who put forward a special framework useful for enhancing the software service-placement procedure within a hybrid cloud environment. They aimed to improve latency without increasing costs. Similarly, Bjorkqvist et al. [27] undertook to analyze the total performance and cost of running services on hybrid clouds. Similarly, Charrada and Tata [7] advanced an FBR (Forward Backward Refinement) algorithm useful for placing service-based applications across hybrid Clouds. The algorithm is designed to help minimize the costs generated by deploying cloud-related services. In turn, Abbes et al. [6] put forward a novel hybrid cloud-related placement optimization approach that rests heavily on the principle of GA. The idea of this optimization process lies in minimizing the public cloud service deployment cost. The experimental results turned out to reveal well the proposed approach's remarkable outperformance in respect of the FBR algorithm [7] in terms of cost. In turn, Cerroni et al. [28] proposed a hybrid cloud placement algorithm, dubbed Business-Driven Management as a Service Plus (BDMaaS+), constructed over the genetic Algorithm fundamentals and principles.

Concerning Rahimi et al. [29], they devised a Simulated Annealing (SA) optimization, useful for implementation with mobile applications. It takes the form of a set of services modelled for execution either via user devices or via a cloud framework (whether local or public). As regards Bittencourt et al. [30,31], they proposed HCOC (Hybrid Cloud Optimized Cost) resource scheduling mechanism to solve the problem of resource requirement. The HCOC enables executing workflows within a specified budget and execution time frame, using DDVR (Dynamic Deployment Virtual Resource) to enhance the resource search process (by enabling adequate resource retrieval based on QoS requirements). As to Bossche et al. [32], they suggested a selection of scheduling algorithms relying on the Earliest Deadline First (EDF) principle. Their design was intended to address the cost optimization problem associated with deadline-constrained applications while accounting for the data constraints, data locality and inaccuracies considerations in task runtime estimates. In turn, Unuvar et al. [33] introduced a hybrid cloud placement algorithm, constructed on the Biased Importance Sampling (BSA) technique, which relies heavily on the application structure to allocate multiple VMs.

### *3.3. Service placement in federated cloud*

This section is devoted to depicting the most recently conducted works elaborated on investigating the area of the federated cloud environment. Worth mentioning, in this regard, is the work undertaken by Altmann et al. [12], whereby a cost model, specifically designed to fit for application with federated hybrid clouds, was devised. Their advanced cost minimization algorithm is helpful for cloud service placement implementation, aiming to minimize computational service costs. Concerning Breitgand et al. [34], they suggested addressing the challenge of managing the efficient provisioning of elastic cloud services through a federated approach. Their placement algorithm is aimed at maximizing the provider profit rates while protecting the consumer offered QoS. To this end, a number of integer programming (IP) formulations were advanced to deal with the placement of VM workloads within and across multiple cloud providers jointly collaborating in a federation.

In [2] and [35], the authors attempted to optimize the placement of software components in federated cloud environments, respectively, using a traditional GA and a GA based memetic algorithm with integer vector representation. For Aryal et al. [36], the centralized genetic algorithm was adopted to address the NP-hard problem associated with heterogeneous resource allocation within a Mobile Edge Computing (MEC) system. As to the authors in [36], they considered applying the optimization principle to achieve an effective VM placement and resource utilization on fog nodes for the sake of meeting the application set requirement. Regarding the work of Espling et al. [19], they formulated the service placement problem into an integer linear program. From a graph, a set of constraints has been extracted based on services components. A mathematical model presents the obtained placement schemes. In the same venue, the authors in [37] proposed a formal description of the service deployment problem, focusing on business process models. They took into account security and availability as the tariffs of cloud providers. Regarding the characteristics of cloud federation, the authors considered bandwidth and pricing parameters. A configurable process model is then employed to produce the placement scheme while adjusting the federation requirements of the enterprise and the cloud.

### *3.4. A summary of research works dealing with service placement in the cloud environment*

This section involves a summary depicting a compilation of the significant approaches advanced to treat the main issues relevant to maintaining an effective service placement within a Cloud Computing environment. They are classified into three main categories: private or public cloud, hybrid and federated cloud, as figured in Table 1 below.

Table 1. A summary of the major service placement related works.

| Reference | Cloud type | | | Optimization Technique | Objectives |
|---|---|---|---|---|---|
| | Private/ Public | Hybrid | Federated | | |
| [3] | X | | | ACO | Resource |
| [5], [25] | X | | | PSO | Execution time, SLA |
| [4] | X | | | Meta-heuristic | Resource utilization, performance |
| [13], [21], 22] | X | | | GA | Delay, Migration Cost |
| [14] | X | | | GA | Throughput, Latency, Cost |
| [15] | X | | | PSO | Delay, Cost, Availability, Reliability |
| [16] | X | | | GA | Latency, Cost |
| [17] | X | | | EA, Local Search | Energy, Latency |
| [18] | X | | | SA, GA | Energy, Latency |
| [20] | X | | | k-cut principle | Execution time, Communication cost |
| [23] | X | | | Meta-heuristic | Energy, SLA |
| [24] | X | | | PSO | Execution time, SLA, security |
| [6] | | X | | GA | Cost |
| [7] | | X | | FBR | Cost |
| [26] | | X | | IP | Latency, cost |
| [27] | | X | | IP | Cost, QoS |
| [28] | | X | | BDMaaS+ | Cost, SLA |
| [29] | | X | | SA | Cost, Power, Delay |
| [30],[31] | | X | | HCOC | Cost |
| [32] | | X | | EDF | Cost |
| [33] | | X | | BSA | Cost, QoS |
| [2], [35] | | | X | GA | Cost |
| [12] | | | X | COMBSPO | Cost |
| [19] | | | X | IP | Cost |
| [34] | | | X | IP and Greedy LP Rounding heuristic | Profit, Performance, Energy Consumption |
| [36] | | | X | GA | Cost |
| [37] | | | X | IP | QoS |

As indicated in table 1, the cloud environments associated service placement problem appears to be addressed from different perspectives, concerning the different types of clouds (private or public, hybrid or federated), as well as the various techniques and criteria applied (Response time, Makespan, QoS, etc.) It is, therefore, clear that all the cited approaches dealing with the service placement subject, considered in this context, turn out to focus on a particular aspect or dimension of this issue. However, with regard to our approach, a clear distinction is established between public-based communication and private-based one. For this reason, different approaches [2,28,29,32,33] are being considered for a jointly hybrid cloud-based architecture to reduce the user's investment. The objective lies in minimizing costs by allowing users to decide on which services to opt for. Service access turns out to be transparent while enhancing scalability, reliability, and reducing costs. Noteworthy, however, is that in attempting to optimize the service placement costs, various approaches do not seem to consider communication flow between the different clouds' parameters as significantly involving high costs.

Regarding the works [2,34,36], the authors tend to consider exclusively a single type of communication cost involving a node or service within the cloud. However, as far as our work is concerned, we consider distinguishing between two major communication modes: public communication and hybrid communication. Actually, to the best of our knowledge, this criterion seems to be accounted for only in the [7] and [6] elaborated works. Still, both approaches do not appear to consider the execution time dimension, which stands as a crucial factor in the company's decision process. Accordingly, our major focus of interest is primarily on treating the service placement generated costs (e.g., hosting cost, inter-service communication cost) and the relevant execution time factor.

## 4. AN ENHANCED BINARY PARTICLE SWARM OPTIMIZATION ALGORITHM

This section begins with a thorough depiction of the major conducted BPSO related works before presenting our advanced algorithm.

### *4.1. Literature review*

Initially developed to fit for application in a space of continuous values, the PSO soon began to raise several problematic issues as to discrete-valued spaces, in which the variable domain is finite. In attempting to solve such a problem, Kennedy and Eberhart [38] advanced a discrete binary version of PSO. In their devised model, a particle would decide on the "yes" or "no", "true" or "false" options, etc. These binary values might well stand as a representation of a real value within a binary search space. Hence, each particle turns out to use binary values to represent its current position and the best solution position. Like the continuous PSO version, the velocity vector keeps being updated, while the major difference lies in the particle's velocities, which are rather defined in terms of probabilities that take on one or zero. According to this probability, the velocity vector must be exclusively restricted within the range of [0,1]. Hence, the sigmoid function figuring in equation (8) turns out to be the fittest for application for each of its values.

In what follows, we will present the BPSO associated equations. Updating a particle's velocity is executed using the following equations:

$$v_{i+1}^{d} = w \times v_i^d + c_1 r_1 \times (pbest_i^d - x_i^d) + c_2 r_2 \times (gbest^d - x_i^d) \quad (6)$$

$$\begin{aligned} & if \;\; v_{i+1}^{d} > V_{HIGH} \;\; then \\ & \quad v_{i+1}^{d} = V_{HIGH} \\ & else \; if \;\; v_{i+1}^{d} < V_{LOW} \;\; then \\ & \quad v_{i+1}^{d} = V_{LOW} \end{aligned} \quad (7)$$

where:

$V_{LOW}$ = designates low velocity.

$V_{HIGH}$ = denotes high velocity.

As for the particle's position updating process, it applies the following equations:

$$Sig(v_i^d) = \frac{1}{1 + e^{-v_i^d}} \quad (8)$$

$$
\begin{aligned}
&if \;\; Sig(v_i^d) > r_3 \; then \\
&x_{i+1}^d = 1 \\
&else \\
&x_{i+1}^d = 0
\end{aligned}
\tag{9}
$$

where:

$w$ = designates inertial weight;

$v_i^d$ = represents velocity for particle *i* at dimension *d;*

$c_1$ = denotes the acceleration constant;

$r_1$ = is a random value;

$x_i^d$ = represents the position for particle *i* at dimension *d*;

$pbest_i^d$ = is the best previous position of the *i*th particle at dimension *d*;

$c_2$ = *is* acceleration constant;

$r_2$ = random value;

$gbest^d$ = denotes the best global position of all particles at dimension *d*;

$r_3$ = is a random value.

However, worth highlighting that the BPSO algorithm is not without any weaknesses, particularly those relating to local minimum, premature convergence, and poor convergence performance. To remedy these pitfalls, several researchers have suggested modifying the BPSO, which has been subject to intense criticism. In this regard,

Murtza et al. [39] propose an Integer PSO (IPSO). Their parameters take integer values, with the particles updating for an upcoming iteration being subject to probabilistic updating with some probability.

As to Miao et al. [40], they put forward a discrete PSO, which stores reasonable solutions in an external archive to be utilised when updating the particles' best personal positions. Accordingly, a probability-based PSO discretisation method was suggested to update the velocity and the particles' position. In turn, Aygun et al. [41] advanced a modified binary PSO, whereby the optimal solution is affected not only by the particle as well as the global best solution but also by the best solution of the neighbourhood particles in this iteration. As for Dong and Zhao [42], an improved binary PSO was proposed, which consists of applying the greedy algorithm to each particle's position and the redundancy elimination algorithm to eliminate any redundancy of the particles' positions. In [43], however, the authors suggested a binary PSO, which they dubbed gPSO, that rests on simultaneously applying the GA and PSO, implementing the GA operators to boost the PSO.

Hence, it is clear that each of these works turns out to provide a suggested modification of the BPSO based on the nature of the application and its specificity. With respect to our application case, however, we consider putting forward a special enhancement method of the standard BPSO by modifying the particle position's updating equation.

*4.2. The proposed E-BPSO equation*

Unlike the discrete BPSO method, we consider putting forward a new algorithm, which we dub E-BPSO (Enhanced Binary PSO), useful for updating each particle's velocity within a continuous space environment. It is worth noting, in this respect, that the BPSO is sensitive to the saturation of the sigmoid function, which occurs whenever the values reached by the velocity appear to be too high or too low. In these cases, the probability of changing the value of the bit approaches zero, thus limiting the exploration process. Indeed, for zero speed, the sigmoid function returns a probability of only 0.5, which means there is a 50% chance that the bit will flip. Therefore, blocking the speed will delay the apparition of sigmoid function saturation.

Noteworthy is that the same velocity update equations (6-7) have been maintained while modifying the particles' position updating equations. The basic BPSO [38] makes use of the velocity equation (8) sigmoid to derive the particle's position, as indicated by equation (9). However, concerning our proposed E-BPSO algorithm, a special equation is implemented to update the particle's position.

Accordingly, equation (10) is to be incorporated in the E-BPSO to substitute the BPSO equations (8-9), such as:

$$\begin{gathered} \text{if } x_i^d + v_{i+1}^d > 0.5 \text{ then} \\ x_{i+1}^d = 1 \\ \text{else} \\ x_{i+1}^d = 0 \end{gathered} \tag{10}$$

The idea of E-BPSO is inspired from the continuous PSO method. As for the particle's position, we consider applying the same PSO particle position update equation (10).

### *4.3. The E-BPSO algorithm*

As we are dealing with a binary environment context (wherein each position may take either the value 1 or 0), we consider setting the value 0.5 as a threshold, whereby one can decide whether the particle's position will take either the value 0 or 1. Regarding our proposed E-BPSO algorithm, therefore, every particle turns out to enclose a set of bits (1 or 0). Each bit should involve a particular position precisely fit for providing a specific service. Accordingly, if the position turns out to be one, the service will then be hosted in the public cloud, and when its value proves to be equal to zero, the service will then be reserved to the private cloud. Accordingly, any population would involve several particles representing a subset of the entire searching space. Concerning the investigated problem, therefore, every particle would be composed of several zeros and ones; thus, a particle might, for instance, look like 0111001011.

Similarly, the suggested E-BPSO particles are activated within a multidimensional environment, where each particle bit bears a proper velocity and position. As highlighted through algorithm 1, the E-BPSO dimension is defined by the number of bits enclosed in a particle, i.e., the number of services involved in the problem. As regards the stopping condition, we opt for a condition that helps in significantly reducing the execution time. In fact, the execution process should stop once the best solution does not mark any improvement following four successive iterations. And in any case, the execution procedure will finish in less than 0.01 seconds. The fitness function of E-BPSO is defined by the equations (1-5). However, the complexity of the algorithm is $O(n^3)$.

Algorithm 1 presents the advanced E-BPSO algorithm.

Algorithm 1: E-BPSO algorithm

```
1   for each particle I
2     for each dimension d
3       Initialize position x_i^d randomly
4       Initialize velocity v_i^d randomly
5     end for
6   end for
7   iteration k=1
8   while maximum time or minimum error criteria are not attained
9   for each particle i
10    Calculate fitness value (F)
11    if F> pbest_i^d in history then
12      pbest_i^d=F
13    end if
14  end for
15  Choose the particle having the best fitness value as the gbest_i^d
16  for each particle i
17    for each dimension d
18      Calculate velocity according to the equation
19      V_i^d(k+1)= w*V_i^d(k)+ c1*r1*(pbest_i^d-X_i^d) + c2*r2*(gbest^d-X_i^d)
20      if V_i^d(k+1)>V_HIGH then
21        V_i^d(k+1)=V_HIGH
22      else if V_i^d(k+1)<V_LOW then
23        V_i^d(k+1)=V_LOW
24      end if
25      Update particle position according to the equation
26      if X_i^d(k)+V_i^d(k)>0.5 then
27        X_i^d(k+1)=1
28      else
29        X_i^d(k+1)=0
30      end if
31    end for
32  end for
33  k=k+1
34  end while
```

The next stage involves evaluating our E-BPSO algorithm through establishing a real-benchmark based comparison with other techniques.

## 5. EXPERIMENTAL EVALUATION

We used a real IBM-based dataset [44] comprising 560 BPMN to assess the advanced algorithm performance. All computation times were achieved via Intel Core i7 CPU 2.4 GHz with RAM 12 Go. A selection of graphs, incorporating between 11 and 20 nodes, was also considered, along with ten randomly selected SBA graphs reflecting the design based service compositions. Some of these graphs are dense, while others are sparse. The density is expressed by equation (11).

$$Density = 100 \times \sum edges \Big/ \sum possible_edges \tag{11}$$

The graph characteristics are displayed on table 3.

**Table 3.** The selected graph characteristics.

| Graphs | Nodes | Edges | Hosting needed | Density |
|---|---|---|---|---|
| G1 | 20 | 19 | 469 | 10% |
| G2 | 17 | 28 | 521 | 20% |
| G3 | 18 | 46 | 418 | 30% |
| G4 | 11 | 22 | 254 | 40% |
| G5 | 16 | 60 | 413 | 50% |
| G6 | 14 | 55 | 332 | 60% |
| G7 | 13 | 55 | 319 | 70% |
| G8 | 19 | 137 | 570 | 80% |
| G9 | 15 | 95 | 363 | 90% |
| G10 | 12 | 66 | 297 | 100% |

The service-based applications are usually depicted in graphs, figuring in a sparse, dense or full form, wherein density represents an essential criterion. By means of illustration, and for an effective assessment of our proposed algorithm, the possible number of edges in G7 is calculated as (13*12)/2=78, where 55 edges have been selected, yielding a density range of 55/78=70%.

As Table 3 indicates, the graphs display varying densities, reflecting the different service nodes' composition. Ten graphs have been selected based on the benchmark, revealing different density rates, ranging from 10% to 100%.

Three of the graphs reflecting our benchmark graphs highlighted density have been selected, namely:

- g1 is a sparse graph, with a number of edges too close to the minimal number of edges;
- g10 is a dense graph, where the number of edges is too close to the maximum number of edges;

- g5 is a graph ranging between sparse and dense that displays a density range of 50%.

It is also worth noting that each of the displayed results corresponding values relates to an average of 10 runs.

The settings parameters used are presented in table 4.

**Table 4.** The setting parameters of E-BPSO

| Name of parameters | Adopted value |
|---|---|
| Swarm size | [50, 200] |
| Max iteration | 500 |
| Problem dimension | [12, 20] |
| C1 | 2.0 |
| C2 | 2.0 |
| W | [0,1] |

This subsection deals with a comparative study involving the various techniques applied in our proposed design. The aim is to highlight the advantages brought about by our advanced E-BPSO algorithm.

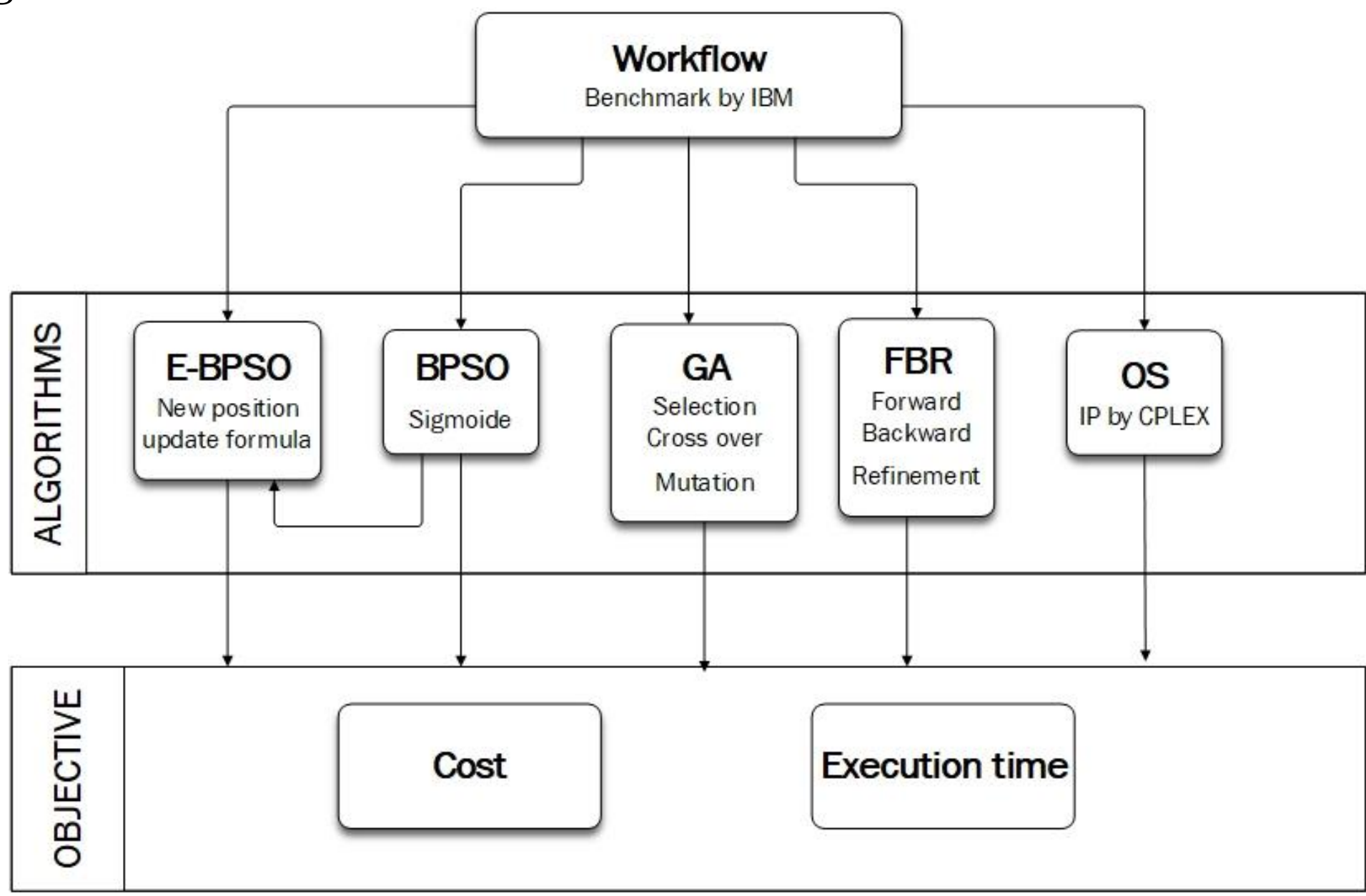

**Figure 4**: Comparative study involving the E-BPSO, BPSO, GA, FBR and OS algorithms.

Figure 4 shows the different techniques implemented in our E-BPSO algorithm compared to the BPSO, GA, FBR, and OS. The E-BPSO turns out to be an enhanced version of the standard BPSO algorithm, designed to boost its performance by substituting the sigmoid function with a new equation (10) within a multidimensional environment. As for the GA rests on a three-operation design, relying on the selection, crossing, and mutation processes. Concerning the FBR, it depends on the forward, backward and refinement procedures. Regarding the OS, it helps maintain the most optimal solution that involves a CPLEX reliant method.

### *5.1. The E-BPSO versus BPSO experiment*

Both E-BPSO and BPSO have been implemented on the three selected graphs: g1, g5 and g10, applying both cost and execution-time measures, with similar parameters being implemented to both algorithms.

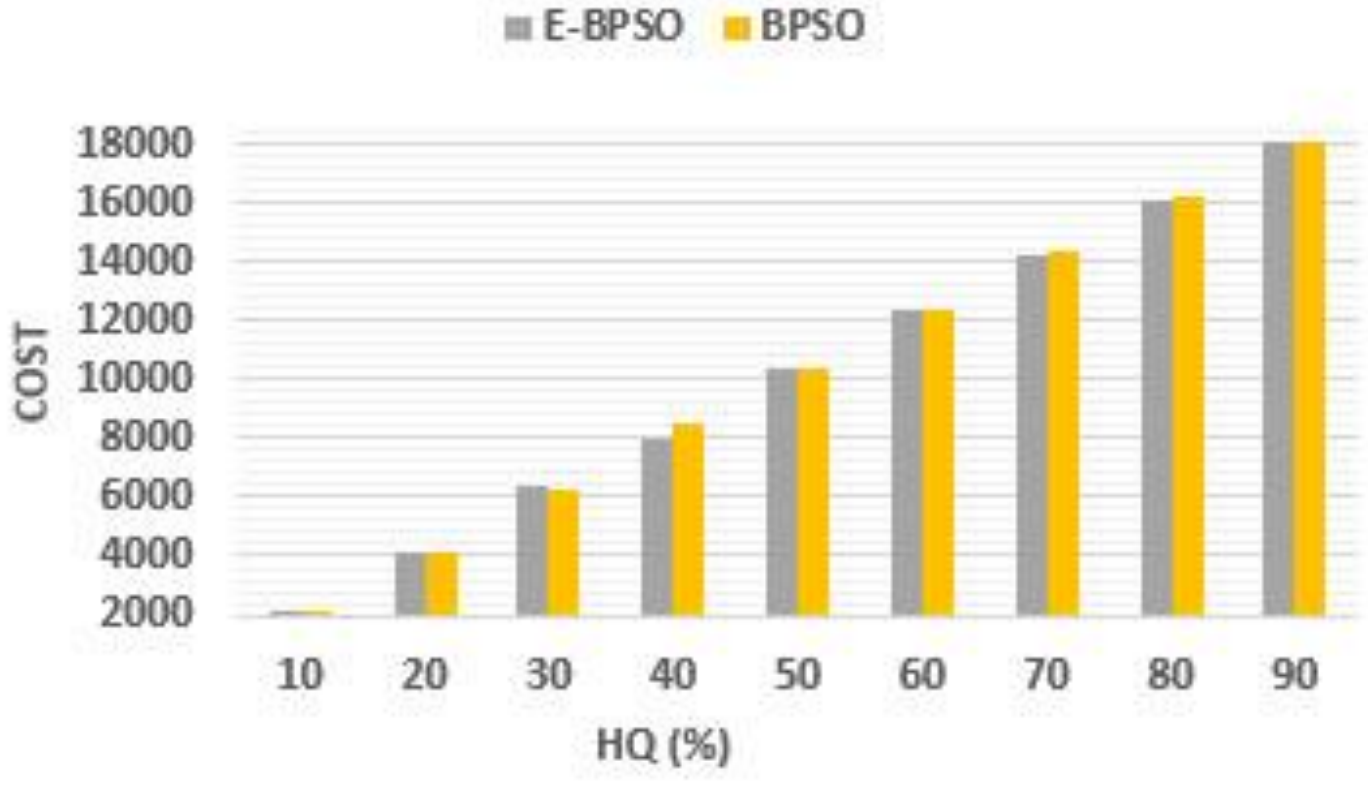

a: Cost based comparison between E-BPSO and BPSO on g1

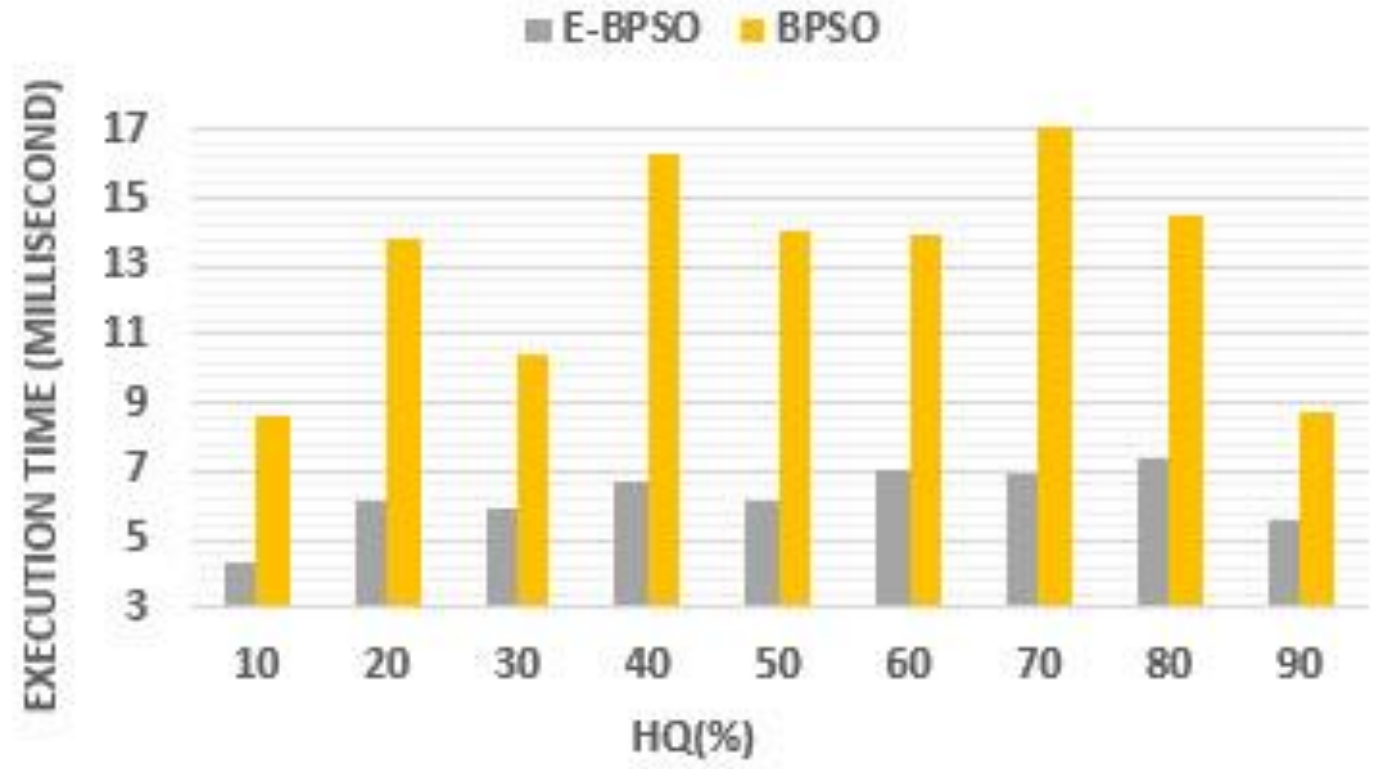

b : Execution-time based comparison between E-BSPO and BPSO on g1

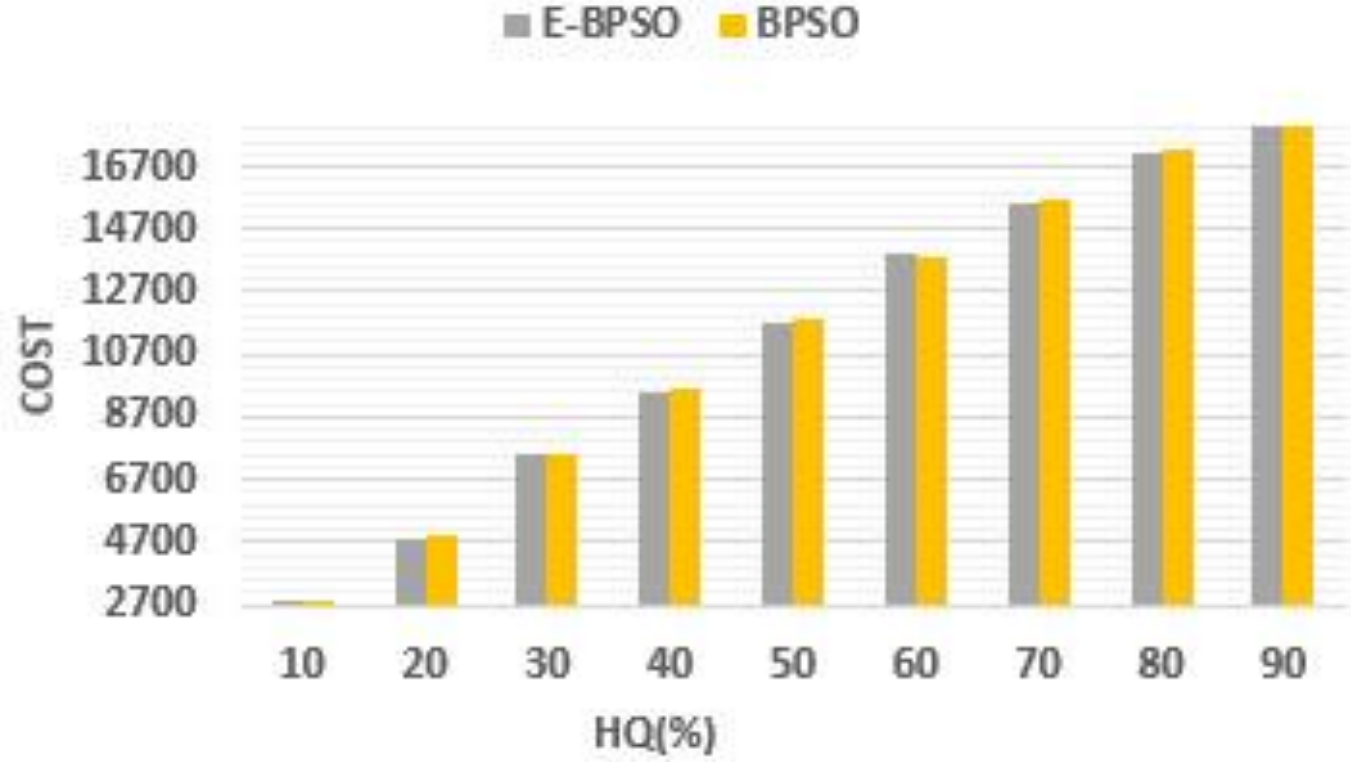

c : Cost based comparison between E-BPSO and BPSO on g5

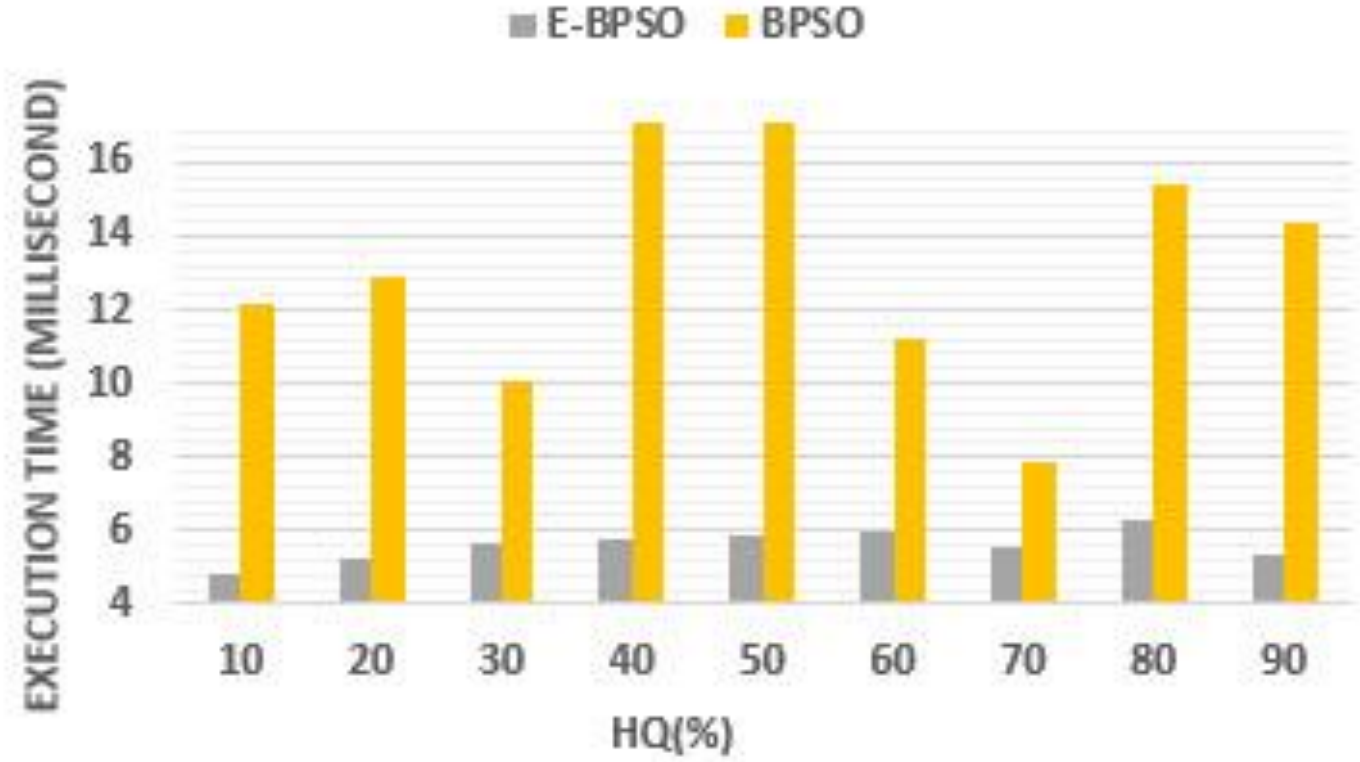

d : Execution-time based comparison between E-BSPO and BPSO on g5

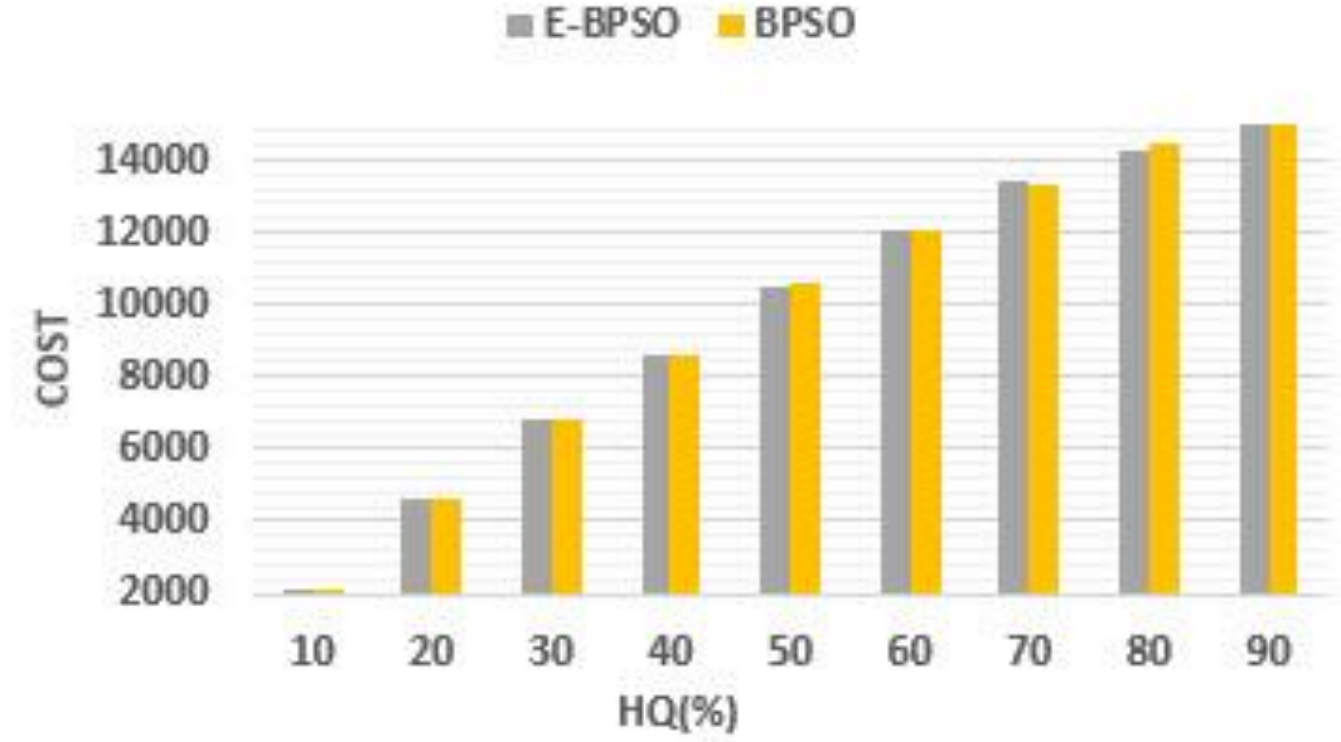

e : Cost based comparison between E-BPSO and BPSO on g10

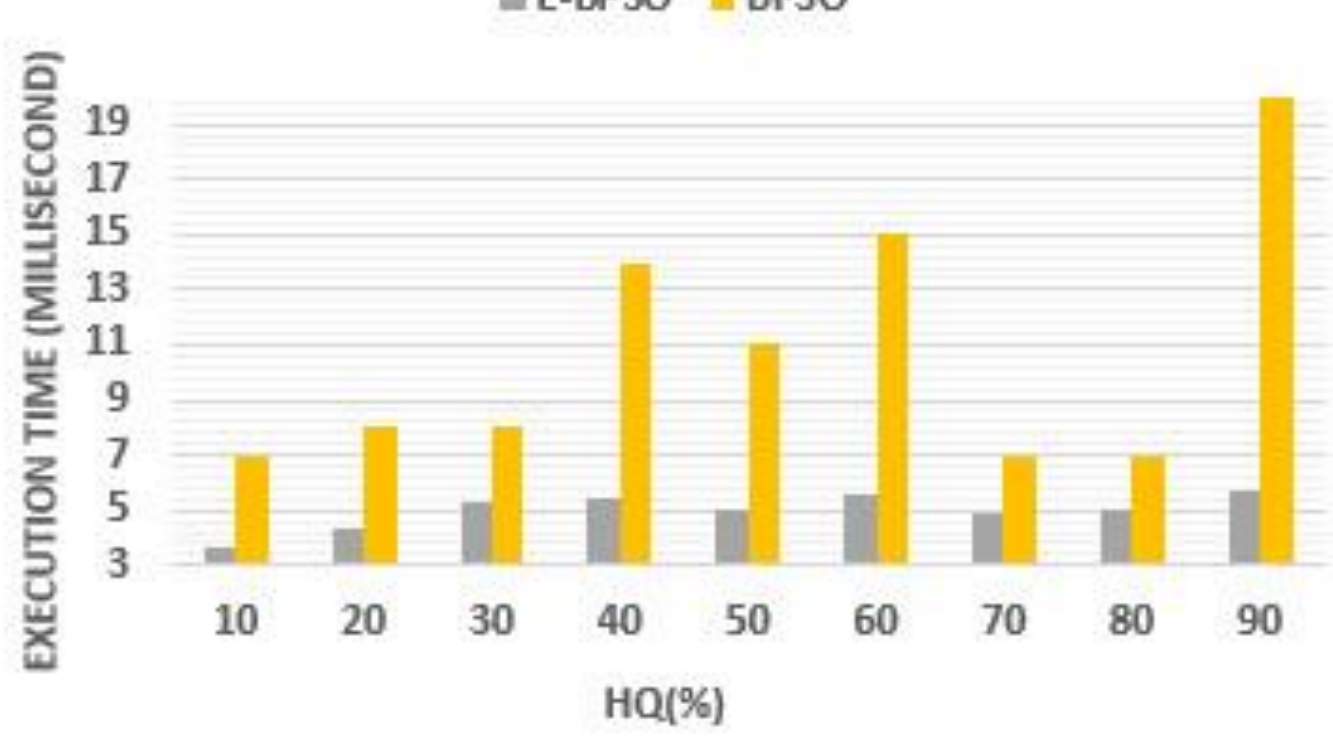

f : Execution-time based comparison between E-BSPO and BPSO on g10

**Figure 5:** Comparison between E-BPSO and BPSO on g1, g5 and g10.

Accordingly, one may well note that in all cases, and regardless of the rate the HQ bears, the E-BPSO appears to achieve an execution time that is noticeably lower than that achieved by the BPSO. Such results could explain the BPSO apparent complexity, which exceeds higher that of the E-BPSO, as highlighted by both algorithms' equations. Regarding the cost related graphs, one can also notice that, in most cases, the E-BPSO turns out to exhibit lower cost rates than the BPSO.

*5.2. Experimental results*

We have used the CPLEX [45] (an IBM developed optimization software package that serves to solve the integer programming problems), to compute the optimal solution (OS). For comparison purposes, the FBR algorithm, initially developed by [7], was applied as an approximate service-placement algorithm, along with the GA, as set up by [6], widely maintained as an effective GA-based placement optimization algorithm.

We administered more than 2150 experiments, and the reached findings turn out to demonstrate that our E-BPSO algorithm helps bring about rather effective results in terms of not only sparse graphs but also dense graphs. Indeed, the entirety of the E-BPSO algorithm achieved results appear to outperform remarkably those attained via the FBR and GA algorithms within the same response time interval. For illustration purposes, some of the achieved results are displayed below. Indeed, as Table 5 indicates, we have studied the different possible scenarios relevant to parameters $\alpha$, $\beta 1$ and $\beta 2$, as delivered by service providers.

**Table 5.** Choice of $\alpha$, $\beta 1$ and $\beta 2$

| Service provider | $\alpha$ | $\beta 1$ | $\beta 2$ | average cost |
|---|---|---|---|---|
| sp1 | 40 | 20 | 10 | 10411 |
| sp2 | 40 | 10 | 20 | 11350 |
| sp3 | 20 | 10 | 40 | 11228 |
| sp4 | 20 | 40 | 10 | 10650 |
| sp5 | 10 | 40 | 20 | 10962 |
| sp6 | 10 | 20 | 40 | 11130 |

As can be noticed, the minimum cost turns to be provided by choice of $\alpha$ =40, $\beta 1$=20 and $\beta 2$=10, with respect to all experiments. It seems logical that the inter-cloud communication cost proves to be more expensive than the intra-cloud communication cost. Therefore, the parameter values have been selected as follows:

- $\alpha$ =40 designates the hosting units' coefficient;
- $\beta 1$ =20 denotes the hybrid communication coefficient;
- $\beta 2$=10 refers to the public communication coefficient.

*5.2.1. Cost*

Among the ten graphs figuring in Table 3, three are going to be considered, specifically: a sparse graph (Figure 6), a complete graph (Figure 8), and a dense graph (Figure 7), in addition to nine different HQ values (ranging from 10% to 90 % of the considered graphs' hosting quantity).

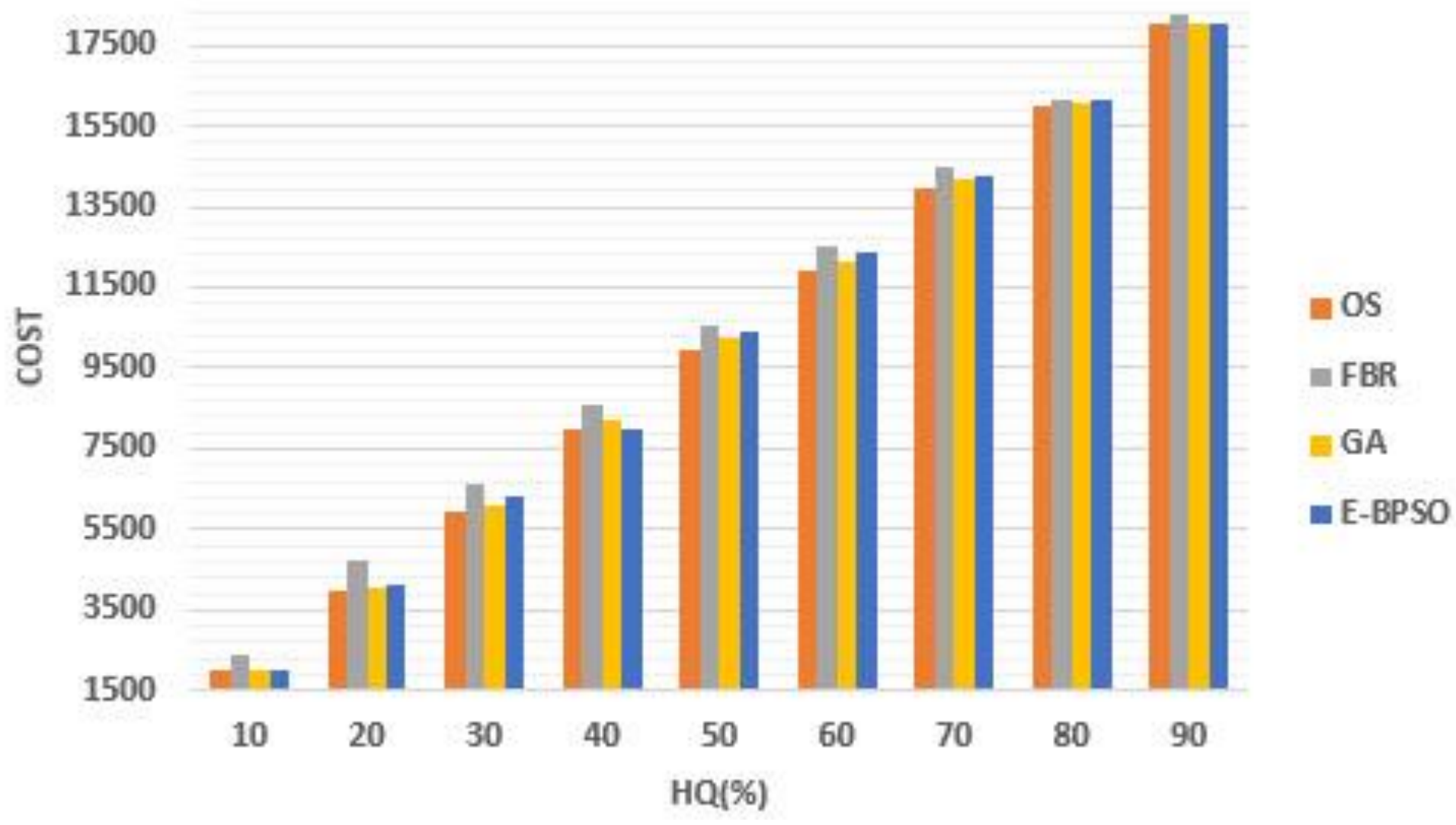

**Figure 6:** Cost comparison between the E-BPSO, GA, FBR and OS in relation to graph G1

Figure 6 depicts the cost values generated by the FBR, GA, BPSO and OS models, relevant to the G1 graph in Table 3, at a density range of 10%.

As can be noted, the E-BPSO cost values appear to be consistently lower than the FBR ones, except for the case where HQ is equal to 80%. The differences in costs recorded between E-BPSO and FBR appear to decrease with increased HQ and vice versa. For instance, at 10% HQ, the recorded difference proves to be very high (16%). These results have their explanation in the sparse graphs displaying a low number of edges (inter-nodal links), and, consequently, the possible solutions turn out to be too low, too.

Noteworthy, also, are the GA achieved results, which slightly outperform those attained by the E-BPSO. This finding can be justified by the GA emitted execution time, which appears to surpass that emitted by the E-BPSO.

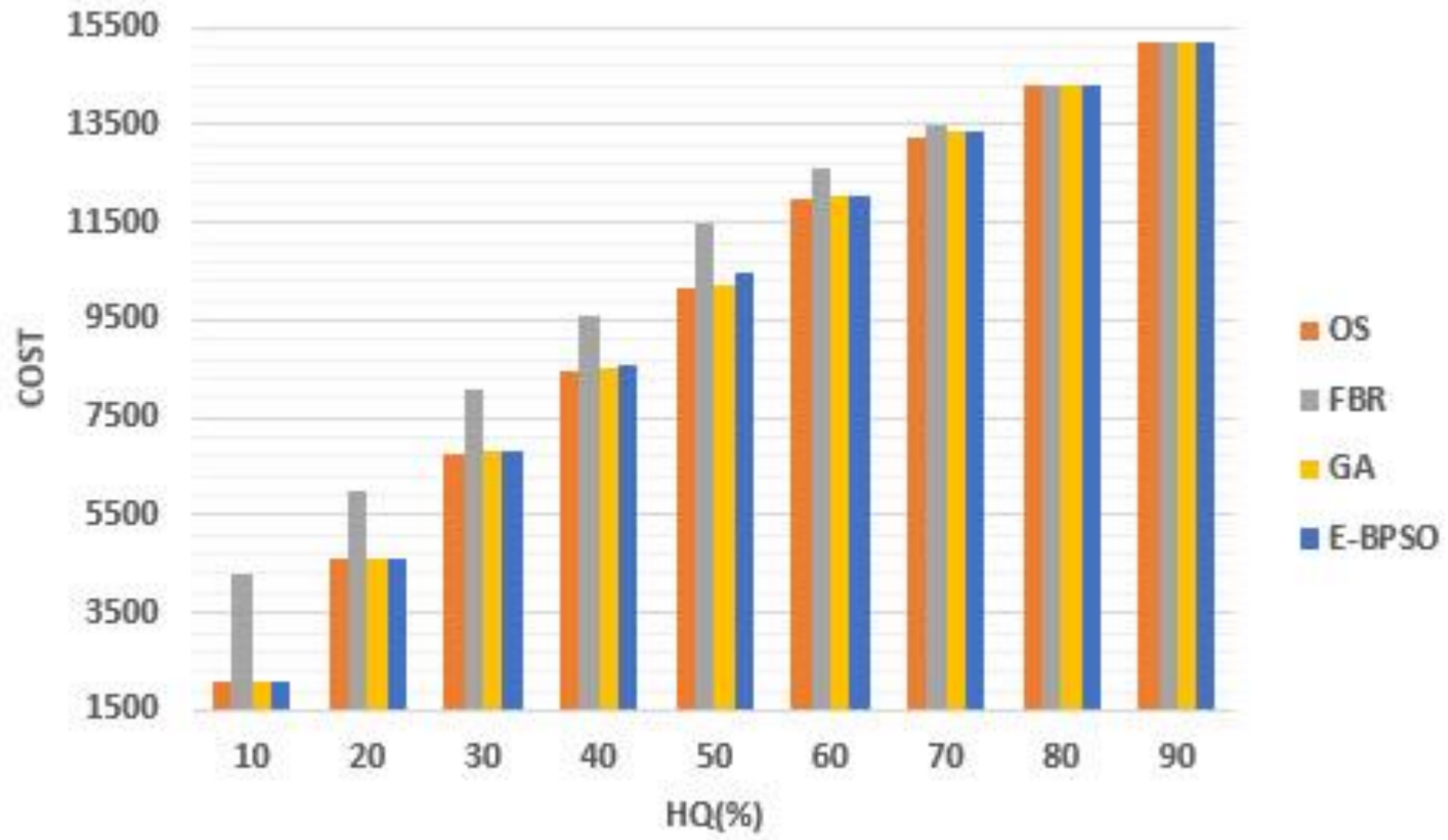

**Figure 7:** Cost comparison between the E-BPSO, GA, FBR and OS in relation to graph G10.

Figure 7 illustrates the cost rates reached on a full graph basis. They highlight well that the E-BPSO achieved costs are too close to the most optimal solutions. Similarly, it appears to obtain effective results, exceeding those achieved by the FBR, concerning all cases. They reveal a distinct rate of around 50%, concerning the case when the HQ rate is 10%. The GA tends to register cost rates that are too close to those obtained by the E-BPSO.

In terms of execution-time cost, the E-BPSO achieved results prove to be rather effective than the FBR attained ones, except for the cases when the HQ exceeds the threshold of 70%, in which the FBR values tend to be equal to, or even slightly exceeding, the E-BPSO values. In practice, however, the HQ should not exceed the threshold of 50% since any company seeking to minimize costs does not often deploy more than 50% of its resource requirements.

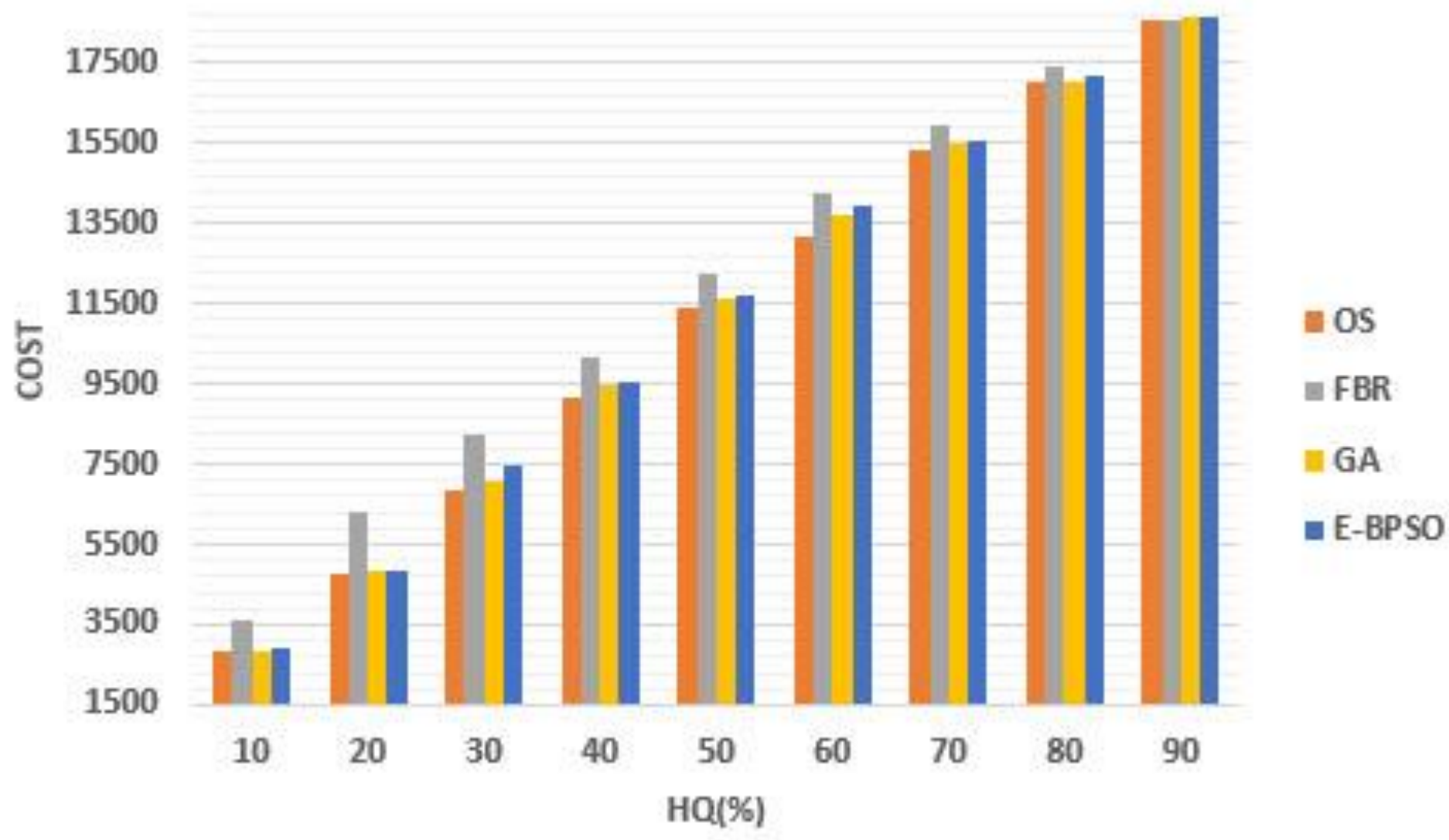

**Figure 8:** Cost comparison between the E-BPSO, GA, FBR and OS in relation to graph G5.

Figure 8 highlights the cost rates attained on a dense graph within a density range of 50%. It indicates well that the E-BPSO tends to perform more effectively than the FBR in most HQ variation cases, except for when HQ is equal to 90%. One could also note that the most optimum cost rate difference achieved between the E-BPSO and FBR is 23%, attained at an HQ level of 20%. Noteworthy is that the GA tends to record cost levels that are somehow too close to, though sometimes slightly higher than, those scored by the E-BPSO.

Analysis of Figures 6-8 reveals that, regardless of graph type, the E-BPSO yields more effective results than the FBR, particularly regarding the dense graphs, where the E-BPSO tends to perform rather efficiently.

*5.2.2. Execution time*

This subsection is focused on examining the second important parameter: execution time, by analyzing the three relevant graphs associated with figures 9, 10 and 11.

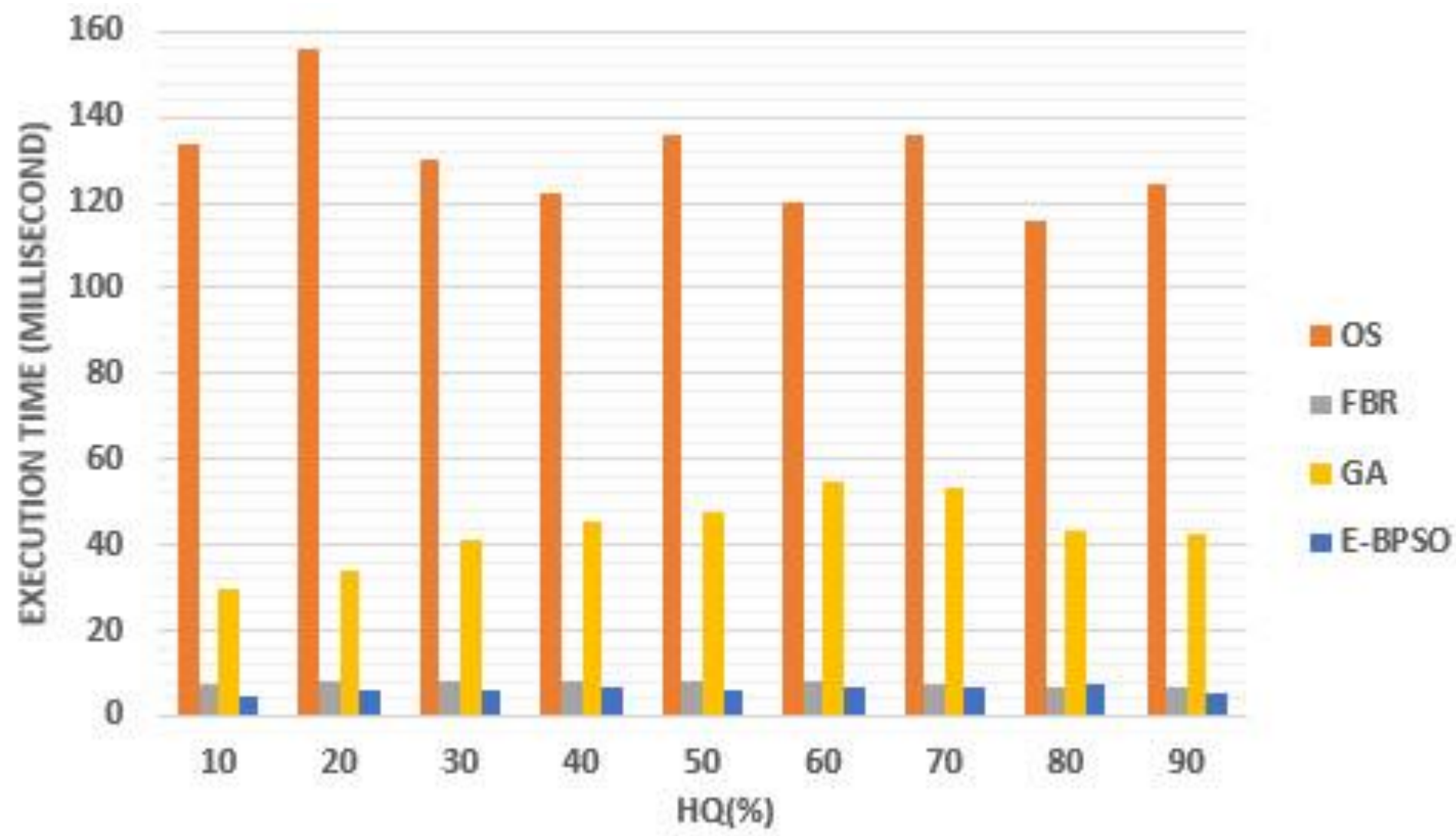

**Figure 9**: Comparison of execution-time performance between the E-BPSO, GA, FBR and OS concerning graph G1.

Figure 9 illustrates the G1 graph execution time parameter regarding the OS, FBR, GA and E-BPSO. As can be noted, the GA is discovered to be more than twice faster than the OS, while the E-BPSO and FBR prove to be more than three times faster than the GA. It has also been revealed that the E-BPSO appears to record an execution time that is noticeably lower than that scored by the FBR, except when the HQ is equal to 80%.

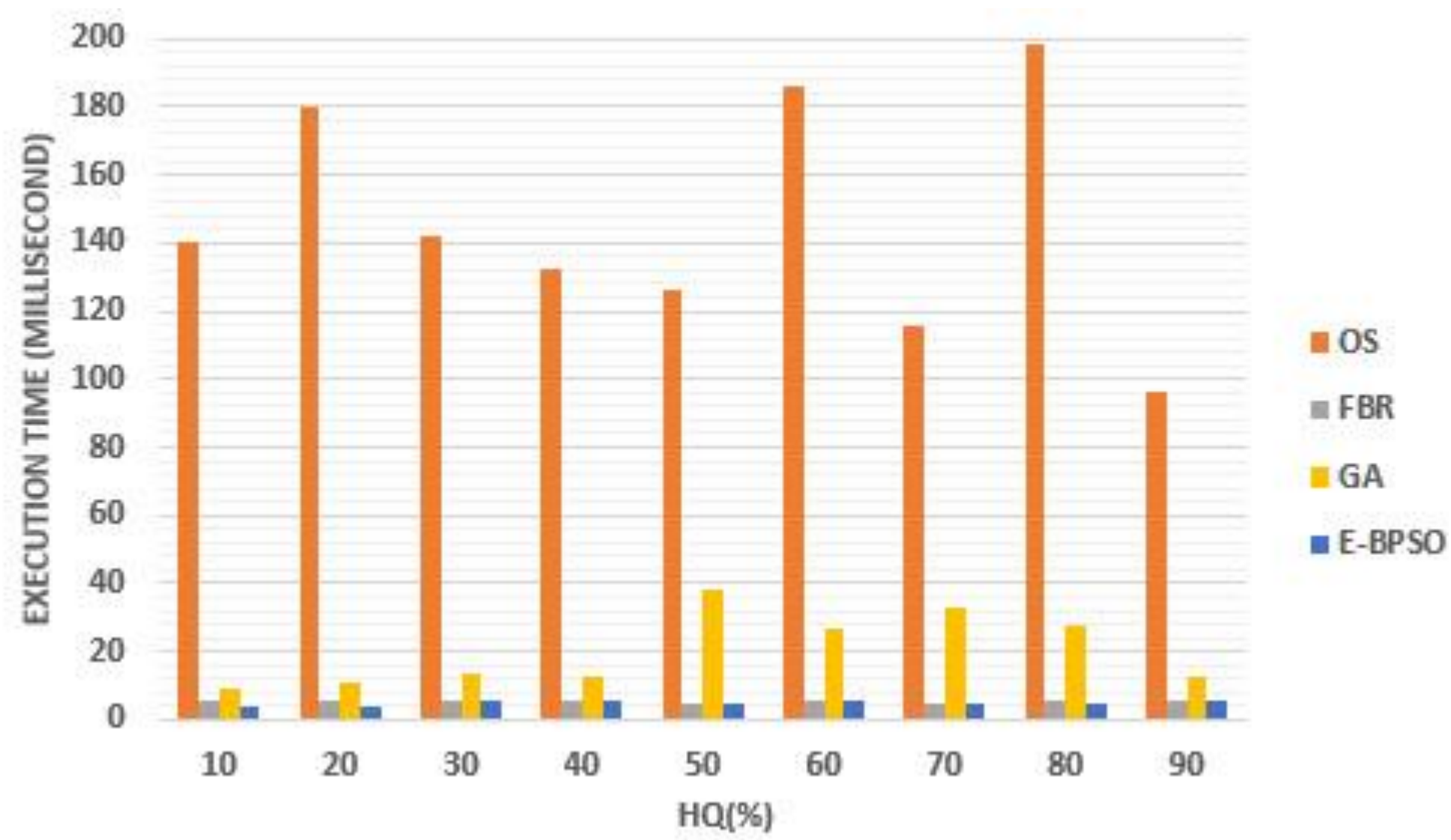

**Figure 10:** Comparison of execution-time performance between the E-BPSO, GA, FBR and OS concerning graph G10.

Accordingly, it has been discovered that the GA appears to record an execution-time performance that is highly effective than that scored by the OS. Inversely, however, both the E-BPSO and FBR tend to score a noticeably better result than that achieved by the GA. Figure 10 also reveals that, for any HQ value, the E-BPSO proves to record the most optimum execution-time level. The execution time difference recorded between both the FBR and E-BPSO turns out to be of the rate of 36%.

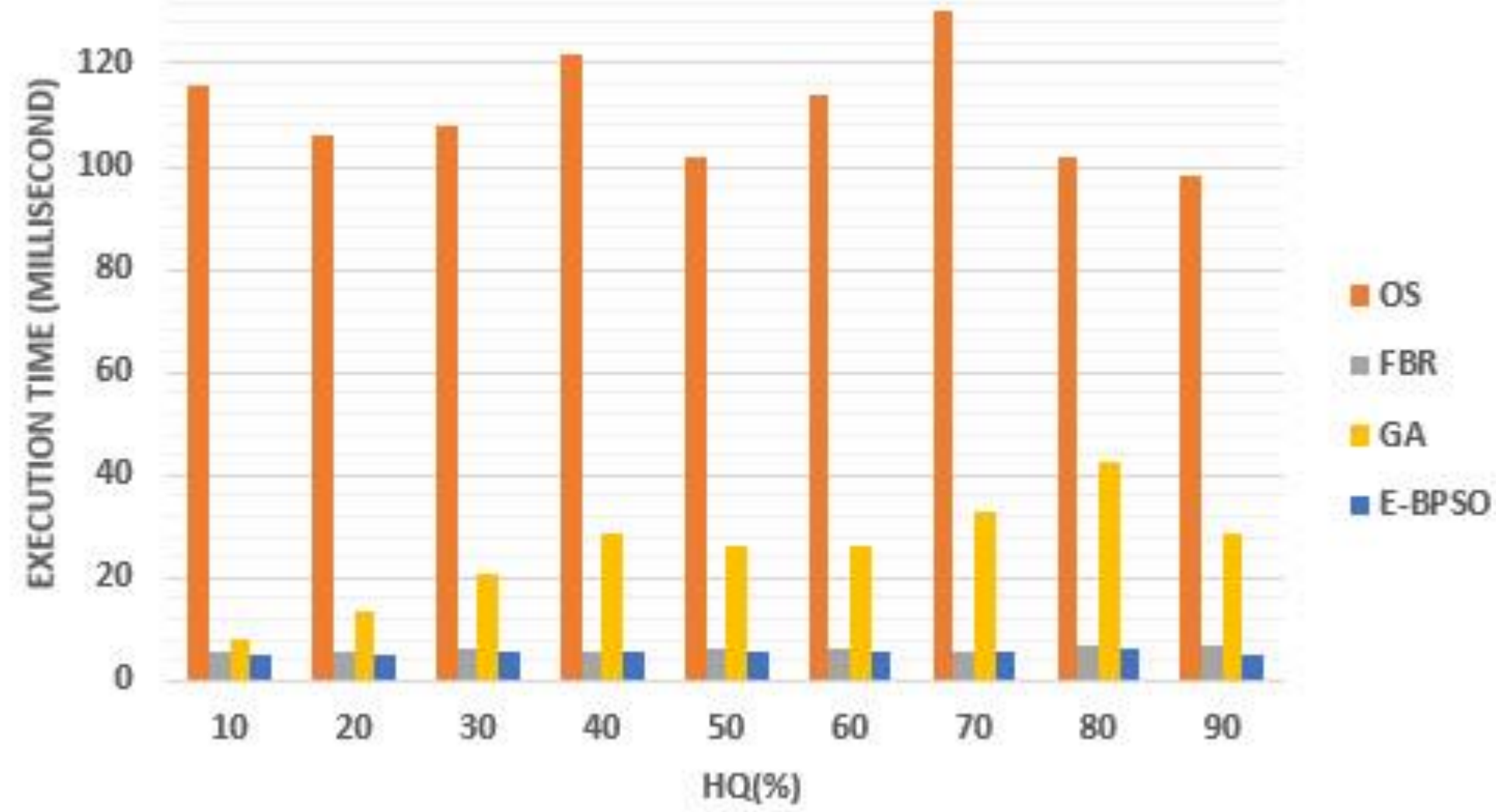

**Figure 11:** Comparison of execution-time performance between the E-BPSO, GA, FBR and OS concerning graph G5.

As Figure 11 indicates, the GA, FBR and E-BPSO tend to register an execution time score that is more than twice as low as that recorded by the OS, while the E-BPSO and FBR appear to record an execution time outperforming that achieved by the GA. Figure 11 also shows that the E-BPSO demonstrates an execution time even lower than that recorded by FBR for all the HQ values. In effect, the most noticeable time difference recorded between the FBR and E-BPSO is of the rate of 24%.

Analysis of Figures 9, 10 and 11 reveals well that, with respect to all graphs, the E-BPSO proves to record the most effective execution time score, as compared to the OS, GA and FBR. This improvement can be justified by the fact that CPLEX pinpointed the most effective solution in terms of cost efficiency while processing all the possible solutions, i.e., from a set of options, despite the noticeable time latency it takes throughout the process. The FBR stands as an iterative algorithm displaying a fixed number of iterations, which entails a minimum execution time to be maintained. Regarding the GA, it demonstrates a more significant time latency in respect of the E-BPSO, owing mainly to the large size population it encloses, whose genetic operations are noticeably consuming in terms of time.

*5.2.3. Comparative study and discussion*

Table 6 illustrates the average percentage improvements brought about by the E-BPSO, in respect of the entirety of the BPSO, GA, FBR, and OS algorithms, in terms of cost and execution time efficiency. These rates refer to the various HQ mean values relevant to each graph. For instance, the E-BPSO provided cost improvement percentage, in relevance to the FBR, is of the rate of 12.69 (graph g1). Note that G1 is a sparse graph, G10 is a dense graph, and G5 is a graph with 50% density. The percentage improvement value is computed via equation (12).

$$\text{Percentage of improvement} = (EBPSO - X) / EBPSO \qquad (12)$$

where EBPSO is the value obtained by the E-BPSO algorithm and X is the value obtained by OS, FBR, GA or BPSO algorithms.

**Table 6.** The E-BPSO provided average improvement percentages in relation to the examined algorithms.

| % of average improvement | Cost | | | | Execution time | | | |
|---|---|---|---|---|---|---|---|---|
| **Graph** | **OS** | **FBR** | **GA** | **BPSO** | **OS** | **FBR** | **GA** | **BPSO** |
| G10 | -2,71 | 4,87 | -0,84 | 0,17 | 95 | 18,59 | 85,51 | 48,32 |
| G5 | -3,06 | 7,57 | -1,32 | 0,24 | 94,9 | 10,13 | 10,1 | 55 |
| G1 | -0,86 | 12,69 | -0,39 | 0,66 | 96,4 | 11,41 | 69,7 | 51,1 |

Analysis of the table also reveals that the proposed E-BPSO algorithm tends to yield rather effective results than the BPSO in terms of cost and execution time. In relation to the GA and OS, however, the E-BPSO appears to exhibit slightly less effective results in terms of cost, within a rate of 2% compared to the GA and a rate of 4% compared to the OS. Yet, in terms of execution time, the E-BPSO proves to record far more highly effective results, highlighting its supremacy over both the OS and GA. This finding can be justified by the fact that even though the OS helps effectively retrieve the most optimal solution, it turns out to generate greater execution time exceeding the rate of 94%. Similarly, the GA also appears to generate an execution time significantly exceeding that registered by the E-BPSO.

Finally, the E-BPSO attained results, achieved in terms of execution time and cost, prove to outperform noticeably those reached via FBR. Only in some cases where the HQ proves to exceed the rate of 70% did the obtained values appear to equal the E-BPSO scored ones, wherein the FBR achieved results appear to be slightly higher. Practically, however, the HQ rate should not exceed the threshold of 50%, as a company seeking to minimize costs should not usually deploy more than 50% of its resource needs.

## 6. CONCLUSION

This paper presented an enhanced algorithm, dubbed E-BPSO, to optimize the SBA placement in a hybrid cloud. It has been designed to reduce the cost of service deployment with a minimal execution time parameter for an efficient selection of the most optimal service placement solution. Thus, the advanced E-BPSO helps resolve the problem of premature convergence and local optima, significantly affecting the service placement application in the hybrid cloud, making it highly distinguishable and relevant to the standard BPSO algorithm.

The reached experimental results reveal that the proposed algorithm E-BPSO appears to display a remarkably effective performance compared to the OS, FBR and GA algorithms in terms of both cost and execution time. It is worth noting that if the number of services is small, the difference between the BPSO and the E-BPSO turns out to be similar in terms of cost and execution time. But, if the number of services is large, the difference in execution time becomes noticeable. However, E-BPSO falls when the number of services is high, or the HQ is greater than 70%.

Our future work will improve and extend the proposed algorithm to be applied to a real workflow that considers new constraints and the dynamic environment.

## ACKNOWLEDGMENT

We deeply acknowledge Taif University for Supporting this study through Taif University Researchers Supporting Project number (TURSP-2020/327), Taif University, Taif, Saudi Arabia.

The present research, contributing in achieving the highlighted promising results, has received funding grant from the Ministry of Higher Education and Scientific Research of Tunisia, under grant agreement number LR11ES48.

## CONFLICT OF INTEREST

We have no Conflict of Interest.

## REFERENCES

[1] Sahni, S., 1974. Computationally related problems. In SIAM Journal on Computing, vol. 3, pp. 262–279.

[2] Foschini, L., Tortonesi., M., 2013. Adaptive and business-driven service placement in federated Cloud computing environments. In IFIP/IEEE International Symposium on Integrated Network Management (IM 2013), IEEE.

[3] Ni, Z.W., Pan, X.F., Wu, Z.J., 2012. An ant colony optimization for the composite saas placement problem in the cloud. In: Applied Mechanics and Materials, vol. 130, pp. 3062–3067. Trans Tech Publ.

[4] Filelis-Papadopoulos, C. K., Endo, P. T., Bendechache, M., Svorobej, S., Giannoutakis, K. M., Gravvanis, G. A., Tzovaras, D., Byrne, J., Lynn, T., 2020. Towards simulation and optimization of cache placement on large virtual content distribution networks. In Journal of Computational Science, Volume 39.

[5] Hajji, M.A., Mezni, H., 2017. A composite particle swarm optimization approach for the composite saas placement in cloud environment. Soft. Comput., pp. 1–21.

[6] Abbes, W., Kechaou, Z., Alimi, A. M., 2016. A New Placement Optimization Approach in Hybrid Cloud Based on Genetic Algorithm. In IEEE International Conference on e-Business Engineering (ICEBE), pp. 226-231.

[7] Ben Charrada, F., Tata, S., 2016. An efficient algorithm for the bursting of service-based applications in hybrid Clouds. In IEEE Transactions on Services Computing, vol. 9, issue 3, pp. 357–367.

[8] Business Process Incubator, April 2021. https://www.businessprocessincubator.com/category/type/templates/

[9] O. M. G. (OMG), Business Process Model and Notation™ (BPMN™) Version 2.0, Object Management Group (OMG), Tech. Rep., jan 2011. http://www.omg.org/spec/BPMN/2.0/

[10] Van den Bossche, R., Vanmechelen, K., Broeckhove, J., 2010. Cost optimal scheduling in hybrid iaas clouds for deadline constrained workloads. In Proceedings of the 2010 IEEE 3rd International Conference on Cloud Computing, ser. CLOUD '10. Washington, DC, USA: IEEE Computer Society, pp. 228–235.

[11] Luong, N. C., Wang, P., Niyato, D., Wen, Y., Han, Z., 2017. Resource Management in Cloud Networking Using Economic Analysis and Pricing Models: A Survey. In IEEE Communications Surveys & Tutorials, vol. 19, issue 2, pp. 954–1001.

[12] Altmann, J., Kashef, M. M., 2014. Cost model based service placement in federated hybrid clouds. In Future Generation Computer Systems, vol. 41, pp. 79–90.

[13] Yusoh, Z., Tang, M., 2010. A penalty-based genetic algorithm for the composite SaaS placement problem in the cloud. In 2010 IEEE Congress on Evolutionary Computation (CEC), pp. 1–8.

[14] Wada, H., Suzuki, J., Yamano, Y., Oba, K., 2012. E3: A multiobjective optimization framework for SLA-aware service composition. IEEE Transactions on Services Computing, vol. 5, no. 3, pp. 358–372.

[15] Li, W., Zhong, Y., Wang, X., Cao, Y., 2013. Resource virtualization and service selection in cloud logistics. Journal of Network and Computer Applications, vol. 36, no. 6, pp. 1696–1704.

[16] Klein, A., Ishikawa, F. Honiden, S., 2014. SanGA: A self-adaptive network-aware approach to service composition. IEEE Transactions on Services Computing, vol. 7, no. 3, pp. 452–464.

[17] Phan, D. H., Suzuki, J., Carroll, R., Balasubramaniam, S., Donnelly, W., Botvich, D., 2012. Evolutionary multiobjective optimization for green clouds. In Proceedings of the 14th Annual Conference Companion on Genetic and Evolutionary Computation, GECCO '12, (New York, NY, USA), pp. 19–26, ACM.

[18] Yao, F., Yao, Y., Xing, L., Chen, H., Lin, Z., Li, T., 2019. An intelligent scheduling algorithm for complex manufacturing system simulation with frequent synchronizations in a cloud environment. Memetic Computing 11, pp. 357–370.

[19] Espling, D., Larsson, L., Li, W., Tordsson, J., Elmroth, E., 2016. Modeling and placement of cloud services with internal structure. IEEE Trans. Cloud Comput. 4(4), pp. 429–439.

[20] Huang, K.-C., Shen, B.-J., 2015. Service deployment strategies for efficient execution of composite saas applications on cloud platform. J. Syst. Softw. 107, pp. 127–141.

[21] Yusoh, Z., Tang, M., 2012. A penalty-based grouping genetic algorithm for multiple composite saas components clustering in cloud. In 2012 IEEE International Conference on Systems, Man, and Cybernetics (SMC), pp. 1396–1401.

[22] Yusoh, Z., Tang, M., 2012. Composite SaaS placement and resource optimization in cloud computing using evolutionary algorithms. In 2012 IEEE 5th International Conference on Cloud Computing (CLOUD), pp. 590–597.

[23] Barthwal, V., Rauthan, M.M.S., 2021. AntPu: a meta-heuristic approach for energy-efficient and SLA aware management of virtual machines in cloud computing. Memetic Computing 13, pp. 91–110.

[24] Mezni, H., Sellami, M., Kouki, J., 2018. Security-aware saas placement using swarm intelligence. J. Softw. Evol. Process, e1932.

[25] Mezni, H., Kouki, J., 2017. A multi-swarm based approach with cooperative learning strategy for composite saas placement. In: Proceedings of the Symposium on Applied Computing, pp. 399–404. ACM.

[26] Kaviani, N., Wohlstadter, E., Lea, R., 2012. Manticore: A framework for partitioning software services for hybrid cloud. In Proceedings of the 2012 IEEE 4th International Conference on Cloud Computing Technology and Science (CloudCom), ser. CLOUDCOM '12. Washington, DC, USA: IEEE Computer Society, pp. 333–340.

[27] Björkqvist, M., Chen, L. Y., Binder, W., 2012. Cost-driven service provisioning in hybrid clouds. In 2012 Fifth IEEE International Conference on Service-Oriented Computing and Applications (SOCA), IEEE.

[28] Cerroni, W., Foschini, L., Grabarnik, G. Ya., Shwartz, L., Tortonesi, M., 2018. Service Placement for Hybrid Clouds Environments based on Realistic Network Measurements. In 14th International Conference on Network and Service Management (CNSM), IEEE.

[29] Rahimi, M., Venkatasubramanian, N., Mehrotra, S., Vasilakos, A., 2012. MAPCloud: Mobile applications on an elastic and scalable 2-tier cloud architecture. In 2012 IEEE Fifth International Conference on Utility and Cloud Computing (UCC), pp. 83–90.

[30] Bittencourt, L. F., Senna, C. R., Madeira, E. R. M., 2010. Scheduling service workflows for cost optimization in hybrid clouds. In 2010 International Conference on Network and Service Management, IEEE.

[31] Bittencourt, L. F., Madeira, E. R. M., 2011. HCOC: a cost optimization algorithm for workflow scheduling in hybrid clouds.

In Journal of Internet Services and Applications, vol. 3, no. 2, pp. 207-227.

[32] Van den Bossche, R., Vanmechelen, K., Broeckhove, J., 2013. Online cost-efficient scheduling of deadline-constrained workloads on hybrid clouds. In Future Generation Computer Systems, vol. 4, no. 29, pp. 973-985.

[33] Unuvar, M., Steinder, M., Tantawi, A. N., 2014. Hybrid cloud placement algorithm. In 2014 IEEE 22nd International Symposium on Modelling, Analysis & Simulation of Computer and Telecommunication Systems, IEEE.

[34] Breitgand, D., Marashini, A., and Tordsson, J., 2011. Policy-driven service placement optimization in federated clouds. In IBM Research Division, Tech. Rep, H-0299.

[35] Grabarnik, G. Y., Shwartz, L., Tortonesi, M., 2014. Business-driven optimization of component placement for complex services in federated Clouds. In Network Operations and Management Symposium (NOMS), IEEE.

[36] Aryal, R. G., Altmann, J., 2018. Dynamic application deployment in federations of clouds and edge resources using a multiobjective optimization AI algorithm. In Third International Conference on Fog and Mobile Edge Computing (FMEC), IEEE.

[37] Rekik, M., Boukadi, K., Assy, N., Gaaloul, W., BenAbdallah, H., 2016. A linear program for optimal configurable business processes deployment into cloud federation. In: 2016 IEEE International Conference on Services Computing (SCC), pp. 34–41. IEEE.

[38] Kennedy, J., Eberhart, R. C., 1997. A discrete binary version of the particle swarm algorithm. IEEE International Conference on Systems, Man, and Cybernetics.

[39] Murtza, S.A., Ahmad A., Qadri, M.Y., Qadri, N.N., Ahmed, J., 2018. Optimizing energy and throughput for mpsocs: an integer

particle swarm optimization approach. J Comput, Vol 100, pp. 227–244.

[40] Miao, Z., Yong, P., Mei, Y., Quanjun, Y., & Xu, X., 2021. A discrete PSO-based static load balancing algorithm for distributed simulations in a cloud environment. Future Generation Computer Systems, Vol. 115, pp. 497–516.

[41] Aygun, B., Kilic, B. G., Arici, N., Cosar, A., Tuncsiper, B., 2021. Application of binary PSO for public cloud resources allocation system of video on demand (VoD) services, Applied Soft Computing, Vol. 99.

[42] Dong, C., Zhao, L., 2019. Sensor network security defense strategy based on attack graph and improved binary PSO, Safety Science, Vol. 117, pp. 81-87.

[43] Ozsoydan, F., B., Baykasoglu, A., 2019. A swarm intelligence-based algorithm for the set-union knapsack problem, Future Generation Computer Systems, Volume 93, pp. 560-569.

[44] Fahland, D., Favre, C., Koehler, J., Lohmann, N., Volzer, H., Wolf, K., 2011. Analysis on demand: Instantaneous soundness checking of industrial business process models. In Data and Knowledge Engineering, vol. 70, issue 5, pp. 448–466.

[45] ILOG SA, ILOG CPLEX 12, User's Manual, 2021. Available: https://www.ibm.com/support/pages/node/134239.